# Understanding the errors of SHAPE-directed RNA structure modeling

*Wipapat Kladwang[1], Christopher C. VanLang[2], Pablo Cordero[3], and Rhiju Das[1,3,4]**

Departments of Biochemistry[1], Chemical Engineering[2], Biomedical Informatics[3], and Physics[4], Stanford University, Stanford CA 94305

* To whom correspondence should be addressed: rhiju@stanford.edu. Phone: (650) 723-5976. Fax: (650) 723-6783.

Running Title: Understanding errors of SHAPE-based RNA modeling

**ABSTRACT.** Single-nucleotide-resolution chemical mapping for structured RNA is being rapidly advanced by new chemistries, faster readouts, and coupling to computational algorithms. Recent tests have shown that selective 2´-hydroxyl acylation by primer extension (SHAPE) can give near-zero error rates (0-2%) in modeling the helices of RNA secondary structure. Here, we benchmark the method on six molecules for which crystallographic data are available: tRNA(phe) and 5S rRNA from *E. coli*; the P4-P6 domain of the *Tetrahymena* group I ribozyme; and ligand-bound domains from riboswitches for adenine, cyclic di-GMP, and glycine. SHAPE-directed modeling of these highly structured RNAs gave an overall false negative rate (FNR) of 17% and a false discovery rate (FDR) of 21%, with at least one helix prediction error in five of the six cases. Extensive variations of data processing, normalization, and modeling parameters did not significantly mitigate modeling errors. Only one varation, filtering out data collected with deoxyinosine triphosphate during primer extension, gave a modest improvement (FNR=12% and FDR=14%). The residual structure modeling errors are explained by insufficient information content of these RNAs' SHAPE data, as evaluated by a nonparametric bootstrapping analysis inspired by approaches in phylogenetic inference. Beyond these benchmark cases, bootstrapping analysis suggests low confidence (<50%) in the majority of helices in a previously proposed SHAPE-directed model for the HIV-1 RNA genome. Thus, SHAPE-directed RNA modeling is not always unambiguous, and helix-by-helix confidence estimates, as described herein, may be critical for interpreting results from this powerful methodology.

The continuing discoveries of new classes of RNA enzymes, switches, and ribonucleoprotein assemblies provide complex challenges for structural and mechanistic dissection [see, e.g., refs. (*1-4*)]. While crystallographic, spectroscopic, and phylogenetic analyses have led to a deeper understanding of several key model systems, the throughput or applicability of these methods is limited, especially for noncoding RNAs that switch between multiple states in their functional cycles (*5-8*). In recent years, several laboratories have revisited a widely applicable chemical approach for attaining nucleotide-resolution RNA structural information, variously called "footprinting" or "chemical structure mapping". Recent advances have included novel chemical modification strategies, faster data analysis software, accelerated readouts via capillary electrophoresis, and multiplexed purification by magnetic beads (*9-14*).

Despite these advances, chemical mapping data are not expected to generally give structure models accurate at nucleotide resolution. To a first approximation, the protection of an RNA nucleotide from chemical modification indicates that it forms some interaction with a partner elsewhere in the system; but these data, by themselves, do not provide enough information to define the interaction partner. Instead, the mapping data can be used to test, refine, or guide structure hypotheses derived from manual inspection or automated algorithms (*15-17*). The accuracy of this approach is necessarily limited by uncertainties in the modeling – including incomplete treatment of non-canonical base pairs, base-backbone interactions, and pseudo-knotted folds (*17*) – and imperfect correlations of chemical modification rates to structural features. Indeed, there are notable historical examples of chemical data giving misleading structural suggestions, including blind modeling work on tRNA (*18, 19*) and 5S ribosomal RNA (*20, 21*).

It was therefore exciting when recent studies of 2´-OH acylation (the SHAPE method) coupled to the *RNAstructure* algorithm reported secondary structure inference with unprecedented sensitivity (98-100% helix recovery) (*17*). The work acknowledged several uncertainties. Measurements were made on ribosomal RNA without protein partners, which may not form the same structures as crystallized protein-bound complexes. For other test cases, the assumed experimental structures were derived from phylogenetic analysis (P546 domain from the *bI3* group I intron), NMR data (HCV IRES), or crystals of constructs with modifications not present in the SHAPE-probed constructs ($tRNA^{Asp}$). A "gold-standard" benchmark of SHAPE-directed secondary structure inference on RNAs with corresponding crystallographic models remains unavailable. We present herein SHAPE data, secondary structure inference, and analysis of systematic and statistical errors for six such RNAs containing a total of 661 nucleotides and 42 helices. Our results provide a rigorous appraisal of the strengths and limitations of this promising chemical/computational technology.

## Experimental Procedures

### *Preparation of model RNAs*

The DNA templates for each RNA (SI Table S1) consisted of the 20-nucleotide T7 RNA polymerase promoter sequence (TTCTAATACGACTCACTATA) followed by the desired sequence. Double-stranded templates were prepared by PCR assembly of DNA oligomers up to 60 nucleotides in length (IDT, Integrated DNA Technologies, IA) with Phusion DNA polymerase (Finnzymes, MA), and purified with AMPure magnetic beads (Agencourt, Beckman Coulter, CA) following manufacturer's instructions. Sample concentrations were measured based on UV absorbance at 260 nm measured on Nanodrop 100 or 8000 spectrophotometers. Verification of template length was accomplished by electrophoresis of all samples and 10-bp and 20-bp ladder length standards (Fermentas, MD) in 4% agarose gels (containing 0.5 mg/mL ethidium bromide) and 1x TBE (100 mM Tris, 83 mM boric acid, 1 mM disodium EDTA).

*In vitro* RNA transcription reactions were carried out in 40 μL volumes with 10 pmols of DNA template; 20 units T7 RNA polymerase (New England Biolabs, MA); 40 mM Tris-HCl (pH 8.1); 25 mM $MgCl_2$; 2 mM spermidine; 1 mM each ATP, CTP, GTP, and UTP; 4% polyethylene glycol 1200; and 0.01% Triton-X-100. Reactions were incubated at 37 °C for 4 hours and monitored by electrophoresis of all samples along with 100-1000 nucleotide RNA length standards (RiboRuler, Fermentas, MD) in 4% denaturing agarose gels (1.1% formaldehyde; run in 1x TAE, 40 mM Tris, 20 mM acetic acid, 1 mM disodium EDTA), stained with SYBR Green II RNA gel stain (Invitrogen, CA) following manufacturer instructions. RNA samples were purified with MagMax magnetic beads (Ambion, TX), following manufacturer's instructions; and concentrations were measured by absorbance at 260 nm on Nanodrop 100 or 8000 spectrophotometers.

*Chemical probing measurements*

Chemical modification reactions consisted of 1.2 pmols RNA in 20 μL with 50 mM Na-HEPES, pH 8.0, and 10 mM $MgCl_2$ and/or ligand at the desired concentration (see SI Table S1); and 5 μL of SHAPE modification reagent. The modification reagent was 24 mg/ml N-methylisatoic anhydride (NMIA) freshly dissolved in anhydrous DMSO. The reactions were incubated at 24 °C for 15 to 60 minutes, with lower modification times for the longer RNAs to maintain overall modification rates less than 30%. In control reactions (for background measurements), 5 μL of deionized water was added instead of modification reagent, and incubated for the same time. For experiments testing DMSO effects, higher concentrations of NMIA in DMSO were prepared and 2 μL of the modification reagent was added to the 20 μL reaction mixture. Reactions were quenched with a premixed solution of 5 μL 0.5 M Na-MES, pH 6.0; 3 μL of 5 M NaCl, 1.5 μL of oligo-dT beads (poly(A) purist, Ambion, TX), and 0.25 μL of 0.5 mM 5´-rhodamine-green labeled primer (AAAAAAAAAAAAAAAAAAAAGTTGTTGTTGTTGTTTCTTT) complementary to the 3´ end of the RNAs [also used in our previous studies (*13, 14*)], and 0.05 μL of a 0.5 mM Alexa-555-labeled oligonucleotide (used to verify normalization). The reactions were purified by magnetic separation, rinsed with 40 μL of 70% ethanol twice, and allowed to air-dry for 10 minutes while remaining on a 96-post magnetic stand. The magnetic-bead mixtures were resuspended in 2.5 μL of deionized water.

The resulting mixtures of modified RNAs and primers bound to magnetic beads were reverse transcribed by the addition of a pre-mixed solution containing 0.2 μL of SuperScript III (Invitrogen, CA), 1.0 μL of 5x SuperScript First Strand buffer (Invitrogen, CA), 0.4 μL of 10 mM each dNTPs [dATP, dCTP, and dTTP; and either dGTP or dITP (*22*)], 0.25 μL of 0.1 M DTT, and 0.65 μL water. The reactions (5 μL total) were incubated at 42 °C for 30 minutes. RNA was degraded by the addition of 5 μL of 0.4 M NaOH and incubation at 90 °C for 3 minutes. The solutions were neutralized by the addition of 5 μL of an acid quench (2 volumes 5 M NaCl, 2 volumes 2 M HCl, and 3 volumes of 3 M Na-acetate). Fluorescent DNA products were purified by magnetic bead separation, rinsed twice with 40 μL of 70% ethanol, and air-dried for 5 minutes. The reverse transcription products, along with magnetic beads, were resuspended in 10 μL of a solution containing 0.125 mM Na-EDTA (pH 8.0) and a Texas-Red-labeled reference ladder (whose fluorescence is spectrally separated from the rhodamine-green-labeled products). The products were separated by capillary electrophoresis on an ABI 3100 or ABI 3700 DNA sequencer. Reference ladders were created using an analogous protocol without chemical modification and the addition of, e.g., 2´-3´-dideoxy-TTP in an amount equimolar to dTTP in the reverse transcriptase reaction.

The HiTRACE software (*23, 24*) was used to analyze the electropherograms. Briefly, traces were aligned by automatically shifting and scaling the time coordinate, based on cross correlation of the Texas Red reference ladder co-loaded with all samples. Sequence assignments to bands, verified by comparison to sequencing ladders, permitted the automated peak-fitting of the traces to Gaussians.

*Likelihood-based processing of SHAPE data*

Quantified SHAPE data were corrected for attenuation of longer reverse transcriptase products due to chemical modification, normalized, and background-subtracted. Rather than using an approximate exponential correction and background scaling (*25*), we used a likelihood framework to determine the final, corrected SHAPE reactivities (see also (*26*)). Furthermore, a likelihood-derived analysis was implemented to average replicate SHAPE data sets across several experiments. Both of these procedures are described in detail in the SI Methods. The algorithms are available in the functions *overmod_and_background_correct_logL.m* and *get_average_standard_state.m* within the freely available HiTRACE software package (*24*). Final averaged data and errors have been made made publicly available in the Stanford RNA Mapping Database (http://rmdb.stanford.edu). The accession IDs are: TRNAPH_SHP_0001, TRP4P6_SHP_0001, 5SRRNA_SHP_0001, ADDRSW_SHP_0001, CIDGMP_SHP_0001, and GLYCFN_SHP_0001.

*Computational modeling*

The *Fold* executable of the RNAstructure package (v5.3) was used to infer SHAPE-directed secondary structures. The entire RNA sequences (SI Table S1), including added flanking sequences, were used for all calculations. The flag "-T 297.15" set the temperature to match our experimental conditions (24 °C). The flags "–sh", "–sm", and "–si" were used to input the SHAPE data file, slope *m*, and intercept *b*. The latter parameters define the pseudoenergy formula $\Delta G_i = m \log( S_i + 1 ) + b$, where $S_i$ is the SHAPE reactivity. In the RNAstructure implementation, these pseudoenergies are applied to each nucleotide that forms an edge base pair, and doubly applied to each nucleotide that forms an internal base pair. Boltzmann probability calculations used the *partition* executable with the same flags.

Nonparametric bootstrapping analysis was carried out as follows. Given normalized SHAPE data $S_i$ for nucleotides $i = 1, 2, .. N$, a bootstrap replicate was generated by choosing $N$ random indices $i'$ from 1 to N, with replacement (*27, 28*) (i.e., some nucleotide positions are not represented and some are present in multiple copies; for the latter, SHAPE pseudoenergies were scaled proportionally). The resulting data sets $S_{i'}$ contained the same number of data points and carried any systematic errors present in the original data set. Secondary structure models directed by these data were analyzed in MATLAB to assess the frequency of each base pair arising in the replicates; the maximum bootstrap value across the base pairs of each helix was taken as the boostrap value for the helix. The bootstrapping analysis is being made available on an automated server at: http://rmdb.stanford.edu/structureserver.

Additional calculations were carried out with the fold() routine of the ViennaRNA package (version 1.8.4; equivalent to the 'RNAfold' command-lines)(*29*) extended to accept SHAPE data and calculate pseudoenergies with the same formula used in RNAstructure; calculations were facilitated through Python bindings available through the software's convenient SWIG (Simplified Wrapper and Interface Generator) interface. Secondary structure figures were prepared with VARNA (*30*).

*Assessment of accuracy*

A crystallographic helix was considered correctly recovered if more than 50% of its base pairs were observed in a helix by the computational model. (In practice, 34 of 35 such helices retained all crystallographic base pairs.) Note that, unlike prior work, helix slips of ±1 were not considered correct [i.e., the pairing (i,j) was not allowed to match the pairings (i,j–1) or (i,j+1)].

## Results

*Accuracy of modeling without experimental data*

The benchmark herein (SI Table S1) collects a diverse set of noncoding RNA domains, containing two classic RNA folding model systems, unmodified tRNA$^{phe}$ from *E. coli* (*31*), and the P4-P6 domain of the *Tetrahymena* group I ribozyme (*32*); a functional RNA that has been a frequent test case for modeling algorithms, the *E. coli* 5S ribosomal RNA (*15, 16, 20, 21*); and three ligand-bound domains from bacterial riboswitches for adenine, cyclic di-GMP, and glycine (*33-39*). For the last RNA (glycine riboswitch from *F. nucleatum*), crystallographic data was not available at the time of modeling but released at the time of manuscript submission; it served as a blind test within our benchmark.

As a control, we first applied the *RNAstructure* (*15, 16*) algorithm *Fold* without any experimental data to the benchmark set (SI Fig. S1). Here and below, we discuss modeling errors in terms of false negative rate (FNR; fraction of crystallographic helices that were missed) and false discovery rate (FDR; fraction of predicted helices that were incorrect). The values are summarized, along with the related statistics of sensitivity and positive predicted value, in Table 1. To highlight features of the RNAs' global folds, we present results in terms of helices rather than individual base pairs. For completeness, FNR, FDR, sensitivity, and positive predictive values at the base-pair level are also compiled in SI Table S2.

Without any data, the *RNAstructure* algorithm missed 16 of 42 helices, giving an FNR of 16/42 = 38%. The models mispredicted an additional 21 helices, giving an FDR of 21/(26 + 21) = 45% (Table 1). These error rates are significantly worse than their ideal values (0%), and confirm the known inaccuracy of current secondary structure prediction methods without experimental guidance [see, e.g., (*16, 17*)].

*Accuracy of modeling with SHAPE data*

We then acquired SHAPE data for each RNA in 50 mM Na-HEPES, pH 8.0, 10 mM $MgCl_2$, and saturating concentrations of ligand (for the three riboswitch domains), using the modification reagent N-methylisatoic anhydride (NMIA). Data quantitation for each RNA involved correction for attenuation of long products, background subtraction, and averaging of 12 to 28 replicates (SI Table S1) guided by a likelihood framework (Methods). The data were in excellent agreement with the expected structures [Fig.1 and Fig. 2 (left panels)]. Strong SHAPE reactivities occur mainly at nucleotides that are outside Watson-Crick helices observed in crystallographic models. Based on prior work (*17*), we expected that inclusion of these data as a pseudo-energy term in the *RNAstructure* algorithm would substantially improve the accuracy of computational models, with helix-level FNR as low as 0-2 %. The improvement was indeed significant, but not to the expected extent (Fig. 2, right panels; Table 1). The FNR decreased from 38% to 17% (missing 7 of 42 helices), and the FDR decreased from 45% to 21% (misprediction of 9 helices). In five of the six RNAs, the calculations failed to recover all the crystallographic helices.

*Evaluating sources of systematic error*

The results above give a somewhat less optimistic picture of SHAPE-directed modeling than previously published measurements (*17*). The differences between SHAPE benchmarks can be most simply ascribed to different test RNAs. Nevertheless, we investigated several other possible systematic explanations for the error rates (FNR and FDR of 17% and 21%, respectively) in our test set. First, we used herein a more stringent evaluation scheme to define helix recovery than previous work (*15-17*), which permitted helix register slips by ±1 (see Methods). Using those less stringent criteria gave similar FNR and FDR of 14% and 18%, respectively. Second, we checked for experimental artifacts. Filtering out nucleotides whose SHAPE pseudoenergy errors exceeded 0.4 kcal/mol gave similar FNR and FDR (14% and 18%; Table 2). Third, to test the quality of our lab's experimental procedures and data processing, we carried out SHAPE measurements on an RNA with a previously published SHAPE-directed model, the hepatitis C virus internal ribosomal entry site domain II. The resulting secondary structure (SI Fig. S2) agreed with prior independent work (*17*). Fourth, primer extension with dNTPs containing dITP instead of dGTP, reduces errors in quantitating 'compressed' bands near G nucleotides (*14, 22, 40*), but gives added variance at C nucleotides due to reverse transcriptase pausing [SI Fig. S3 and (*14*)]. Using only data collected with dGTP gave helix-level FNR and FDR of 12% and 14%, respectively (Table 2) – an improvement, but still higher than values of 0-2% achieved for previous test RNAs. The FNR and FDR increased when we used only data collected with dITP (26% and 28%). Fifth, as an additional check on experimental artifacts, we acquired SHAPE data for all the RNAs with the newly developed 2′-OH acylating reagent 1-methyl-7-nitroisatoic anhydride (1M7) (*41*); the FNR and FDR for models based on these data were identical to the measurements with the more widely used NMIA (Table 2).

Sixth, model accuracy might be unduly sensitive to the highest or lowest reactivities in the SHAPE data. However, capping 'outliers' (see SI Methods); changing the cutoffs for capping; removing outliers; only including high-reactivity data; and excluding SHAPE data for nucleotides near the 5′ and 3′ ends of the RNA did not improve the accuracy (Table 2). Seventh, the pseudo-energy for base-pairing is derived from SHAPE data by a logarithmic formula [$\Delta G = m \log(1.0 + \text{SHAPE}) + b$]. Optimizing the parameters *m* and *b* did not affect FNR and improved FDR only slightly (from 21% to 18%; Table 2). Eighth, choices in normalizing SHAPE data can affect the modeling; but varying the normalization by factors between 0.5-fold to 2-fold did not significantly improve the accuracy (Table 2). Ninth, we explored whether energy inaccuracies stem from *RNAstructure*'s thermodynamic parameters, SHAPE data, or both. Comparing energies of crystallographic vs. model structures indicated that both thermodynamic and SHAPE energies are imbalanced to favor incorrect models (by averages of 1.7 and 1.3 kcal/mol, respectively; SI Table S3). Additionally, shifting the Boltzmann weight balances by raising the modeling temperature from 24 °C to 37 °C did not change the error rates (Table 2). Tenth, we additionally tested for algorithm biases by recomputing models in *ViennaRNA* (*29*) rather than *RNAstructure*, but, overall, the FNR and FDR both increased (to 26% and 28%; Table 2).

*Evidence against crystal/solution-structure discrepancies*

Having found no straightforward explanation for SHAPE-directed modeling errors from systematic errors in experimental data acquisition, data processing, or modeling protocols, we investigated whether there might be differences between these RNA's secondary structures in available crystals and in our experimental solution conditions, as occurred in prior work on extracted ribosomal RNA (*17*). Several lines of evidence disfavor this hypothesis in our cases. For tRNA (phe), the P4-P6 domain, the 5S rRNA, and the purine and c-di-GMP riboswitch, independent crystallographic models of several variants indicate that the RNAs' secondary structures agree with phylogenetic analysis and are furthermore robust to different conditions, binding partners, and crystallographic contexts (SI Table S1). In addition, while flanking sequences added to constructs (SI Table S1) might disrupt the target domains, we designed these sequences to avoid such pairings, and checked this lack of pairings by calculations with and without SHAPE data (SI Fig. S1 & Fig. 2).

Misfolding to kinetically trapped secondary or tertiary structures could lead to differences in solution chemical mapping data compared to those expected from crystallographic structures. To test this possibility, we acquired data for the RNAs after incubating them in 10 mM Na-MES, pH 6.0 and 10 mM $MgCl_2$ for 30 minutes ('refolding' conditions developed for large ribozymes (*42, 43*)); the resulting reactivities were indistinguishable from RNAs without the refolding treatment (see, e.g., SI Fig. S3 for tRNA data). Similarly, we tested for adverse effects of dimethyl sulfoxide (DMSO, used to solubilize the SHAPE reagent) (*44*) by repeating measurements in lower DMSO conditions (10% vs. 25% DMSO); SHAPE data were indistinguishable in the two conditions (SI Fig. S3 gives tRNA data).

In addition to these results disfavoring differences in crystal/solution structures, our solution measurements gave positive evidence for the RNAs folding into the correct tertiary conformations. The P4-P6 domain and the 5S rRNA gave changes in their metal core and loop E regions, respectively, upon $Mg^{2+}$ addition, as expected from prior biophysical analysis [e.g, (*45-48*)]; and the three riboswitches gave SHAPE changes with and without their ligands (SI Fig. S4). Most strongly, we have subjected each of these RNAs to the mutate-and-map method, a two-dimensional extension of chemical mapping (*13, 14*), and observed near-complete recovery of the crystallographic helices [98% sensitivity; (*49*)], indicating that the dominant solution structure matches the structure determined by crystallography.

*Assessing information content and confidence by bootstrapping*

A final explanation for the errors of SHAPE-directed structure models could be that the experimental data have insufficient information content to define the secondary structure. That is, the data, while accurately reflecting each RNA's solution conformation, are also consistent with non-native secondary structures with similar calculated energy. Indeed, the minimum energy model can be highly sensitive to small changes in the SHAPE data (see tRNA example in SI Fig. S5a-c); and, in some cases, the incorrect lowest-energy SHAPE-directed model is within 1 kcal/mol of the crystallographic structure (see $tRNA^{phe}$ and the cyclic-di-GMP riboswitch; SI Table S3). Unfortunately, quantitatively interpreting energy differences between models (as well as partition-function-based base pair probabilities, which are skewed to high values; see SI Fig. S5a) is currently complicated by the non-physical nature of the SHAPE pseudoenergies. For example, a useful confidence value should be a good approximation to the actual modeling accuracy. In contrast, the mean base pair probability value over all predicted helices is 88%, suggesting a false discovery rate of 100% – 88% = 12%, substantially underestimating the actual error rate of 21%.

We therefore estimated the helix-by-helix confidence of SHAPE directed models through a nonparametric bootstrapping procedure, inspired by techniques developed to evaluate phylogenetic trees from multiple sequence alignments (*27, 28, 50*). We generated 400 mock replicates of each data set by resampling with replacement the SHAPE data for individual residues; generating secondary structure models directed by these mock data sets; and evaluating the frequency with which each predicted helix appeared in these replicates (SI Fig. S5b and percentage values in Fig. 2). One quarter of the modeled helices (11 of 44) appeared with bootstrap values under 55%, suggesting insufficient information to confidently determine their structure; 7 of these 11 helices were indeed incorrect. Encouragingly, the 33 helices with bootstrap values above 55% included only two errors, of which one was a single-nucleotide register shift. Further, these bootstrap values are robust to small changes in the SHAPE data (see tRNA example in SI Figs. S6d and e). Finally, the overall mean of the helix bootstrap values was 77%. This result predicts a false discovery rate of $100\% - 77\% = 23\%$, in accord with the actual rate of 21%. Bootstrap analysis therefore appears to be well-suited for evaluating confidence in SHAPE-directed models.

*Bootstrap analysis of an independent test case: the HIV-1 genome model*

As a final demonstration of the utility of bootstrapping confidence estimation, we investigated the information content of an external data set. Recent application of the SHAPE method to the 9173-nucleotide RNA genome extracted from the NL4-3 HIV-1 virion gave a secondary structure hypothesis containing 429 helices (*51*), and the quantitated SHAPE reactivity data have been published. Employing these data and previously used modeling constraints (including division of the modeled genome into five separated domains), the current version of *RNAstructure* (5.3) largely recovers the prior working model (Table S4). Furthermore, bootstrapping revealed additional useful information. Several of the model regions, including the 57-nucleotide 5′ TAR element, two helices with lengths greater than 10 bps in the *gag-pol* region, and the signal-peptide stem at the 5′ end of *gp120*, have bootstrap values above 95% and are thus highly confident. Overall, however, 236 of 429 helices in the prior SHAPE-directed model have bootstrap confidence estimates lower than 50%. (If base pairs across the five assumed domains are permitted, more helices are found with such low bootstrap values.) The bootstrap value averaged over all predicted helices is 49%; excluding 59 stems in the prior model that are not recovered with the current version of *RNAstructure* gives a similar value of 55%. These results suggest that much of the HIV-1 secondary structure remains uncertain, even in regions that are strongly protected from SHAPE modification (SI Fig. S7). These low-confidence regions either form single structures that are poorly constrained by the SHAPE data or interconvert between multiple well-formed structures in solution. A tabulation of the helix-by-helix confidence estimates in SI Table S4 should help guide further dissection of these uncertain regions by other chemical and structural approaches.

## Discussion

With recent experimental and computational accelerations, nucleotide-resolution chemical mapping permits the characterization of non-coding RNAs at an unprecedented rate. Nevertheless, the resulting data are not always sufficient to determine the molecule's secondary structure, especially if additional tertiary interactions are present. The helix-level error rates found in this study of six highly structured RNAs (false negative rate and false discovery rate of 17% and 20%, respectively) are significantly better than models generated without data (38% and 45%, respectively), but higher than for prior SHAPE-modeling test cases (FNR of 0-2%). The modeling inaccuracy found herein is similar to error rates (FNR of ~24%) found in benchmarks with other chemical modifiers including dimethyl sulfate, kethoxal, and carbodiimide (*16*), albeit on different RNAs and with different modeling protocols. Side-by-side tests on the same models RNAs will be necessary to rigorously compare conventional chemical approaches with SHAPE-based methods.

As with all structure characterization methods, SHAPE-directed models cannot be considered "determined structures" but instead are useful hypotheses – especially if accompanied by confidence estimates. This work proposes a bootstrapping analysis for SHAPE-directed modeling that provides such confidence values for novel RNAs. In addition to giving correct predictions for helix accuracy in six crystallized RNAs, bootstrapping analysis of the HIV-1 RNA genome finds numerous regions with high uncertainty in the RNA's current SHAPE-directed working model. More information-rich multidimensional methods, such as NMR and the mutate-and-map chemical approaches (*13, 14*), should be able to test these predictions and, more generally, help attain accurate models of non-coding RNAs.

**ACKNOWLEDGMENT**. We thank authors of *RNAstructure* and *ViennaRNA* for making source code freely available; M. Elazar and J. Glenn for the gift of 1M7; and J. Lucks, D. Mathews, K. Weeks, and the Das lab for manuscript comments. Work was supported by the Burroughs-Wellcome Foundation (CASI to RD), NIH (T32 HG000044 to CVL), and a Stanford Graduate Fellowship (to PC).

**SUPPORTING INFORMATION.** Methods for likelihood-based data processing; four tables with detailed benchmark information and systematic error analyses; and eight supporting figures. This material is available free of charge via the Internet at http://pubs.acs.org. Averaged SHAPE data are available at http://rmdb.stanford.edu.

**Table 1. Accuracy of secondary structure recovery by *RNAstructure* with and without SHAPE data**.

| RNA | Len. | Number of helices[a] | | | | |
|---|---|---|---|---|---|---|
| | | Cryst | RNAstructure | | + SHAPE | |
| | | | TP | FP | TP | FP |
| tRNA$^{phe}$ | 76 | 4 | 2 | 3 | 3 | 1 |
| P4-P6 RNA | 158 | 11 | 10 | 1 | 9 | 1 |
| 5S rRNA | 118 | 7 | 1 | 9 | 6 | 3 |
| Adenine ribosw. | 71 | 3 | 2 | 3 | 3 | 1 |
| c-di-GMP ribosw. | 80 | 8 | 6 | 2 | 6 | 2 |
| Glycine riboswitch | 158 | 9 | 5 | 3 | 8 | 1 |
| Total | 661 | 42 | 26 | 21 | 35 | 9 |
| **False negative rate**[b] | | | 38.1% | | **16.7%** | |
| **False discovery rate**[c] | | | 44.7% | | **20.5%** | |
| **Sensitivity**[d] | | | 61.9% | | 83.3% | |
| **Positive predictive value**[e] | | | 55.3% | | 79.5% | |

[a] Cryst = number of helices in crystallographic model. TP = true positives; FP = false positives.
[b] False negative rate = 1 – TP/Cryst.
[c] False discovery rate = FP/(TP+FP).
[d] Sensitivity = (1 – false negative rate) = TP/Cryst.
[e] Positive predictive value = (1 – false discovery rate) = TP/(TP+FP).

**Table 2. Effects of variations of data processing or modeling on accuracy of SHAPE–directed secondary structure modeling.**

| **Variation in modeling**[a] | **TP**[b] | **FP**[b] | **False negative rate** | **False discovery rate** |
|---|---|---|---|---|
| No SHAPE data (control) | 26 | 21 | 38.1% | 44.7% |
| **SHAPE–directed, default parameters** | **35** | **9** | **16.7%** | **20.5%** |
| Remove residues with high errors[c] | 36 | 8 | 14.3% | 18.2% |
| Use only data collected with dITP during primer extension | 31 | 12 | 26.2% | 27.9% |
| Use only data collected with dGTP during primer extension | 37 | 6 | 11.9% | 14.0% |
| Use 1M7 instead of NMIA reagent | 35 | 9 | 16.7% | 20.5% |
| Cap outliers[d] at cutoff value | 35 | 9 | 16.7% | 20.5% |
| Cap outliers[d] at 2.0 | 35 | 9 | 16.7% | 20.5% |
| Remove additional 5 residues from 5´and 3´ end | 35 | 9 | 16.7% | 20.5% |
| Remove residues with SHAPE < 0.5 | 32 | 14 | 23.8% | 30.4% |
| Optimized *m* and b in pseudoenergy relation[e] | 35 | 8 | 16.7% | 18.6% |
| Adjust normalization 2x | 35 | 8 | 16.7% | 18.6% |
| Adjust normalization 1.5x | 35 | 9 | 16.7% | 20.5% |
| Adjust normalization 0.75x | 34 | 12 | 19.0% | 26.1% |
| Adjust normalization 0.5x | 31 | 14 | 26.2% | 31.1% |
| RNAstructure T=37 °C (not 24 °C) | 35 | 9 | 16.7% | 20.5% |
| ViennaRNA[f] instead of RNAstructure | 32 | 10 | 23.8% | 23.8% |

[a] All variations are described relative to 'default conditions' (in bold) using RNAstructure version 5.3.

[b] The total number of crystallographic helices is 42. TP = true positives; FP = false positives.

[c] Any residues whose estimated measurement error of SHAPE reactivity would give errors of more than ±0.4 kcal/mol if included in a base pair, using the SHAPE pseudoenergy relation.

[d] Outliers were defined as in the normalization procedure: those with values above a cutoff equal to 1.5 times the interquartile range.

[e] Pseudoenergy applied to base–paired nucleotides given by $m \log ( 1.0 + \text{SHAPE} ) + b$. Default parameters in *RNAstructure* are $m = 2.6$ and $b = -0.8$. The combinations of $m$ and $b$ gave the same optimal accuracies for this benchmark were $m = 3.0$ and $b = -0.6$.

[f] ViennnaRNA version 1.8.4, using the default parameter set of Matthews et al. (1999) (*15*).

**Figure 1.** SHAPE reactivities measured at single–nucleotide resolution for six non-coding RNAs of known structure. Black lines mark residues that are paired or unpaired in the crystallographic models with values of 0.0 or 1.0, respectively.

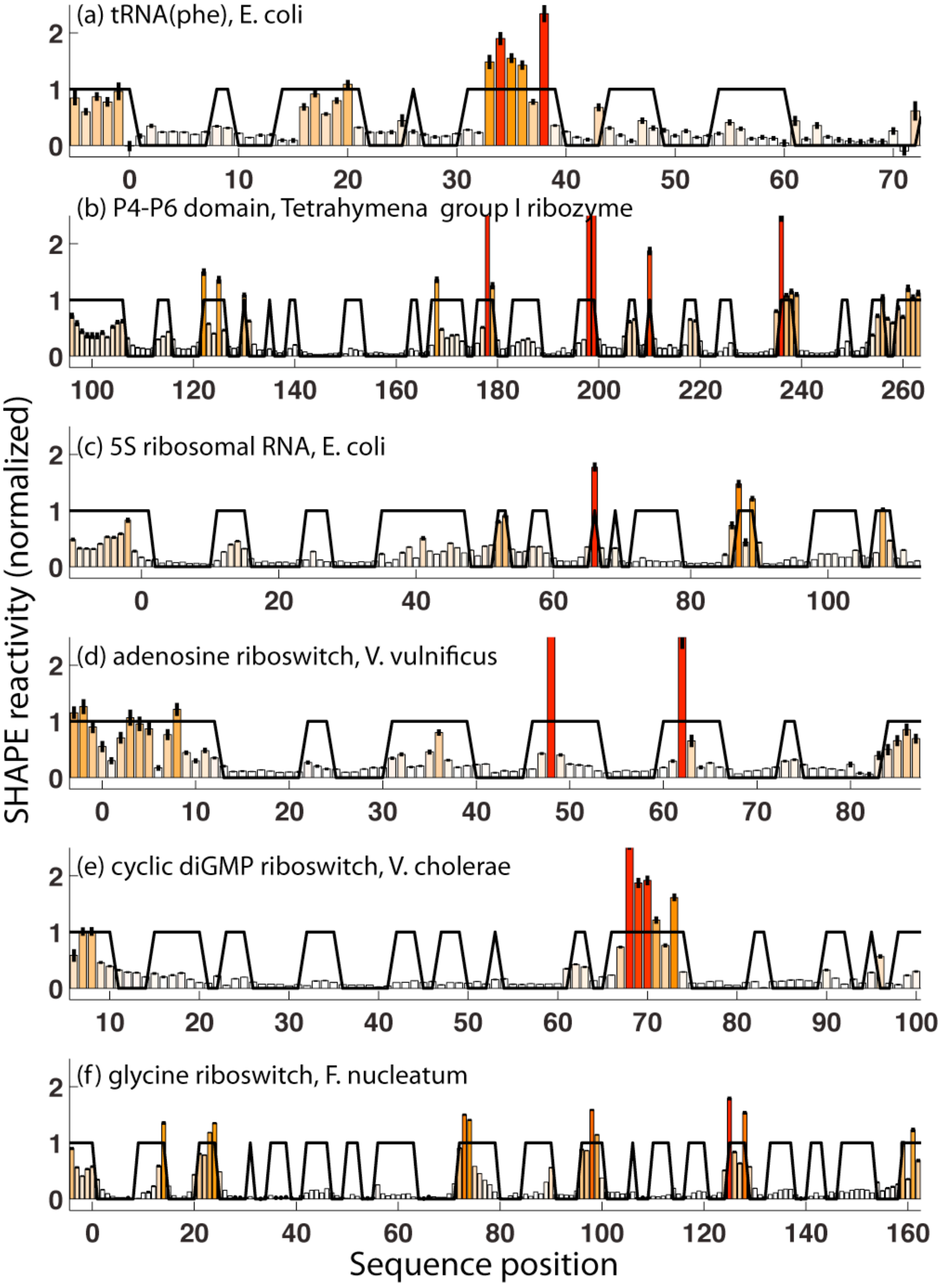

**Figure 2.** Crystallographic (left) and SHAPE-directed (right) secondary structure models for a benchmark of non-coding RNAs. SHAPE reactivities are shown as colors on bases, and match colors in Fig. 1. Cyan lines mark incorrect base pairs; orange lines mark crystallographic base pairs missing in each model; gray lines mark base pairs in regions outside crystallized construct. Helix confidence estimates from bootstrap analyses are given as red percentage values. For clarity, flanking sequences (see SI Table S1) are not shown. Figure is in two parts.

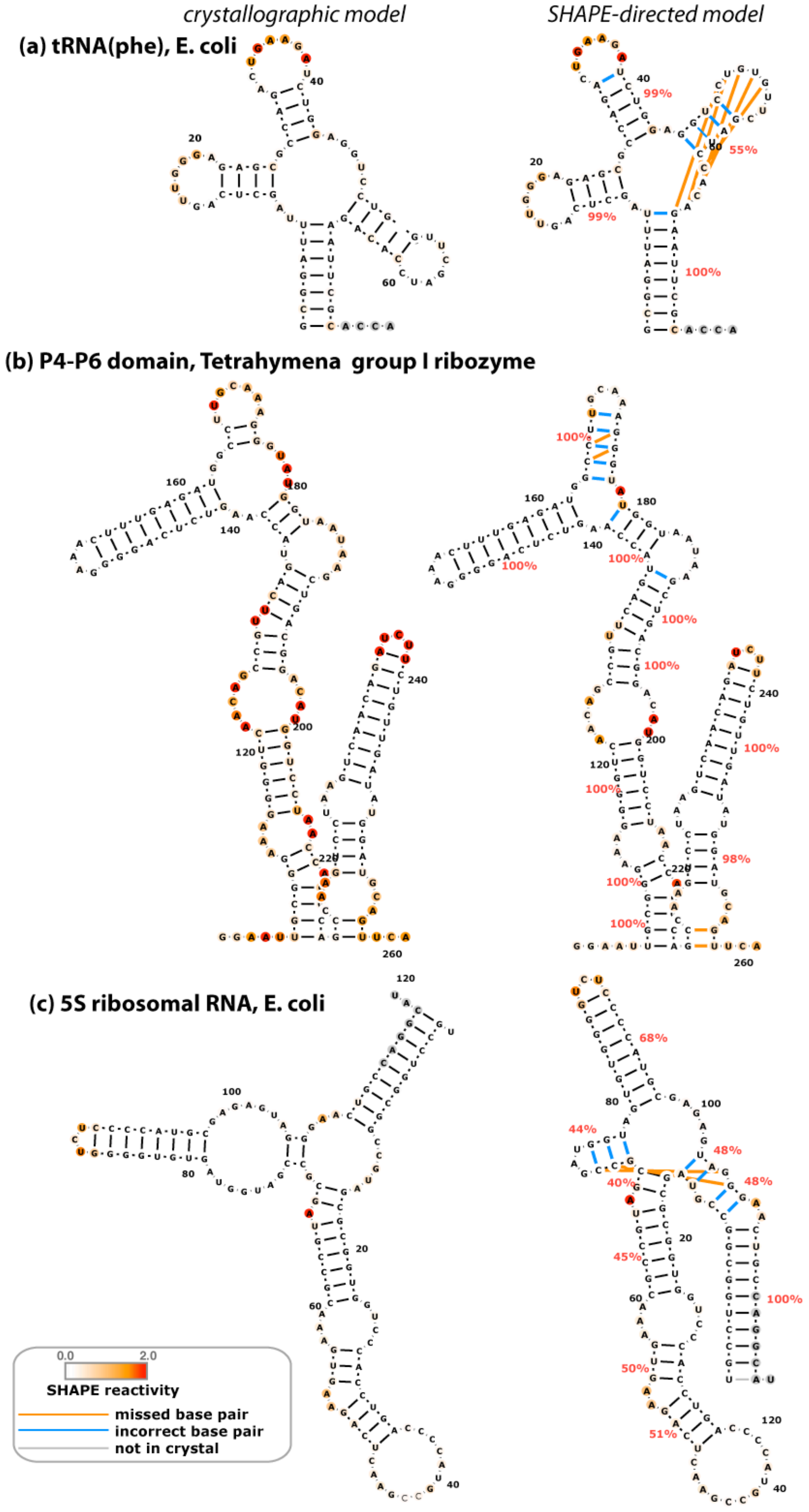

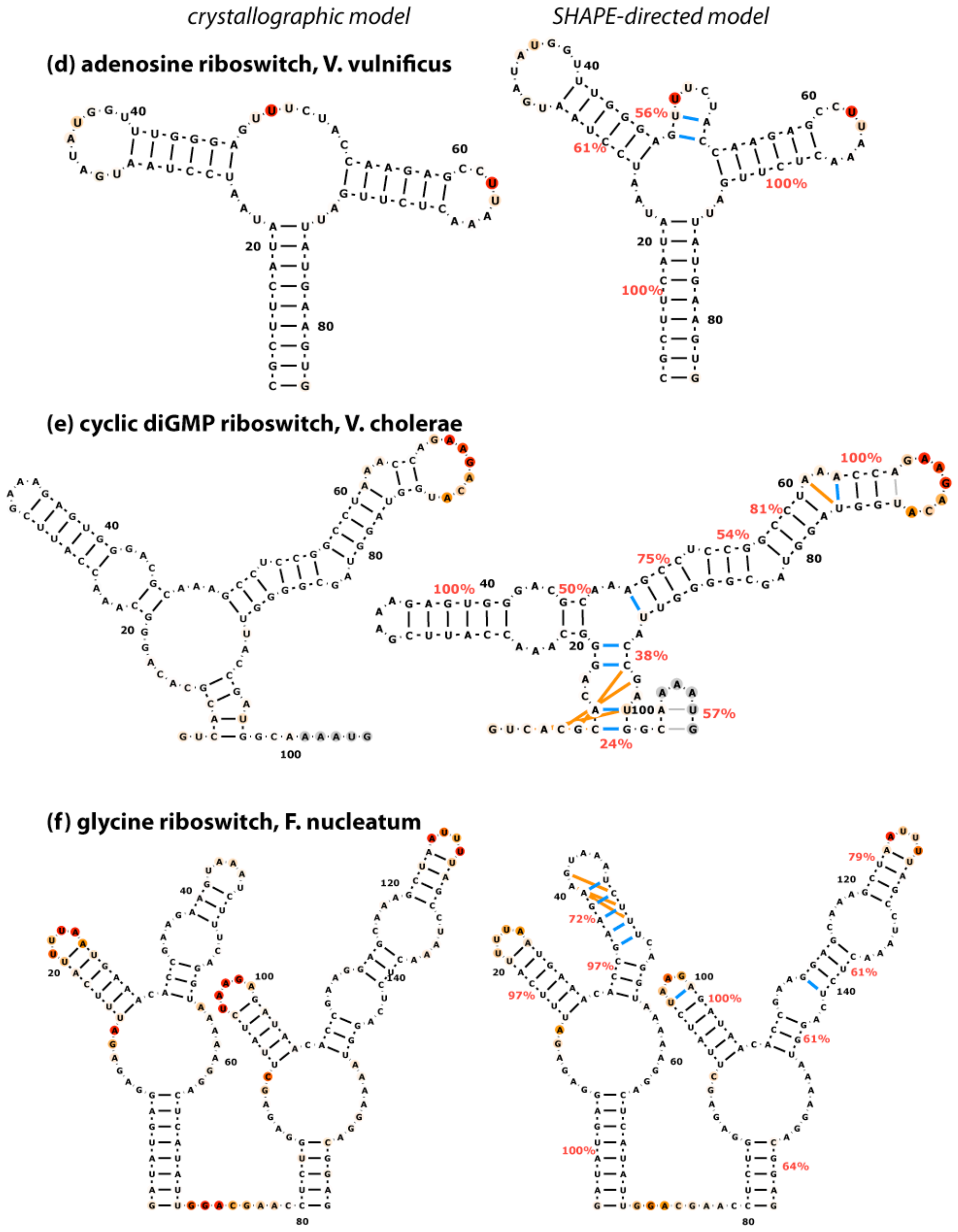
crystallographic model
SHAPE-directed model
(d) adenosine riboswitch, V. vulnificus
(e) cyclic diGMP riboswitch, V. cholerae
(f) glycine riboswitch, F. nucleatum

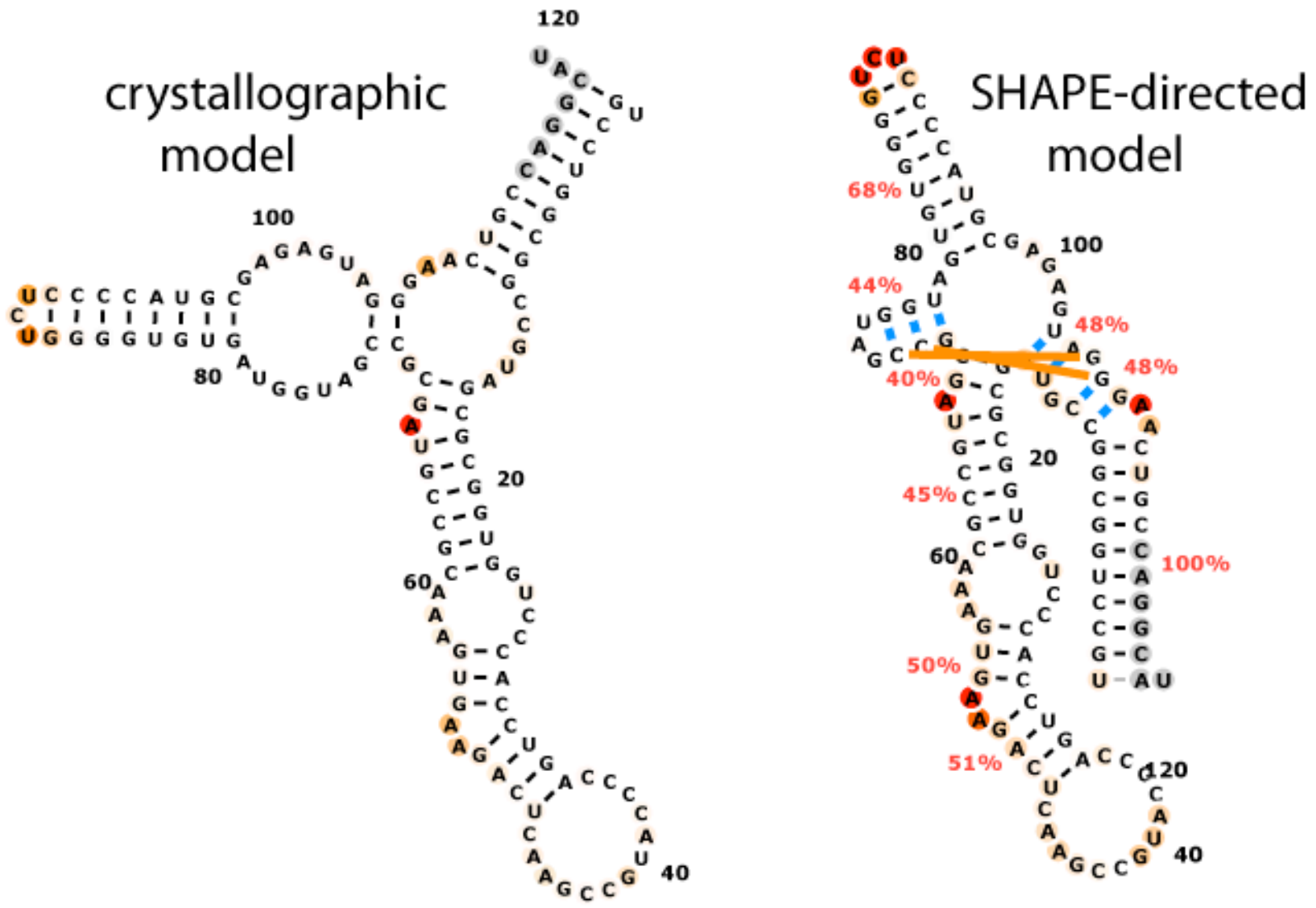

[Table of Contents Figure]

**Supporting Information for:** Understanding the errors of SHAPE-directed RNA structure modeling

Wipapat Kladwang[1], Christopher C. VanLang[2], Pablo Cordero[3], and Rhiju Das[1,3,4]*

*Departments of Biochemistry[1], Chemical Engineering[2], Biomedical Informatics[3], and Physics[4], Stanford University, Stanford CA 94305*

** To whom correspondence should be addressed: rhiju@stanford.edu. Phone: (650) 723-5976. Fax: (650) 723-6783.*

This document contains the following Supporting Information:

## Supporting Methods

*Likelihood-based attenuation correction, background subtraction, and normalization of SHAPE data*

Quantified SHAPE data were corrected for attenuation of longer reverse transcriptase products due to chemical modification, normalized, and background-subtracted. This procedure involved optimization carried out over three parameters, the overall modification rate $\gamma$, a normalization factor for the data $\alpha$, and a scaling for the background $\beta$, defined and estimated as follows. First, let $y_i$ be the observed fraction of products stopping at each nucleotide $i$ = 1, 2, … $N$ in the reverse transcription reaction, ordered so that longer (attenuated) products have larger indices $i$; and define $y_{N+1} = 1 - \sum_{i=1}^{N} y_i$ be the fraction of products that are fully extended. These probabilities are equal to the underlying stopping probabilities $p_i$ times the product of probabilities that the reverse transcriptase has not stopped earlier:

$$y_i = p_i \prod_{j=1}^{i-1} (1 - p_j), \quad (1)$$

which can be inverted by recursively calculating $p_1 = y_1$, then $p_2 = y_2(1-p_1)^{-1}$, then $p_3 = y_3(1-p_1)^{-1}(1-p_2)^{-1}$, etc. For convenience, define $c_i = p_i/y_i$ as an attenuation correction factor. [In the limit that the $y_i$ are approximately constant and much smaller than 1, the solution reduces to an exponential fall-off $c_i^{-1} = \exp[-\langle y_i \rangle i]$, as was effectively assumed in (*1, 2*), but this approximation is unnecessary.]

Fluorescence measurements of reverse transcription products from capillary electrophoresis detectors are in arbitrary units; evaluating (1) requires that the $y_i$ be properly normalized, and this is in principle achieved by the constraint that $\sum_{i=1}^{N} y_i + y_{N+1} = 1$. However, in practice, the intensity of the longest products (e.g., $y_{N+1}$) cannot measured precisely. The strong fluorescence of these fully extended products typically saturates the experimental detector. Thus, the values $y_i$ are defined only up to a proportionality constant γ, i.e., $y_i = \gamma s_i / S_i$ where $s_i$ are the observed fluorescence intensities and $S = \sum_{j=1}^{N_{observed}} s_j$; we need to select amongst $\gamma < 1$.

We estimate γ at the same time as the two other unknown proportionality constants in the data normalization and background subtraction. Let $b_i$ be the quantified band intensities from control measurements (no SHAPE reaction), which we assume require negligible correction from attenuation. The background-subtracted SHAPE reactivities are given in terms of unknown constants α and β by:

$$s_i^{\text{correct}} = \alpha[p_i - \beta b_i] = \alpha[c_i(\gamma)s_i - \beta b_i] \tag{2}$$

We then optimized the log-likelihood function:

$$L = \prod_{i=1}^{n} \alpha c_i(\gamma) \exp\left[F(\alpha[c_i(\gamma)s_i - \beta b_i])\right] \tag{3}$$

Here, $F$ was chosen as a piecewise linear function that (i) gives an exponential distribution of positive SHAPE reactivities [similar to what is empirically observed (SI Fig. S8)], (ii) penalizes negative SHAPE reactivities, and (iii) results in a convex optimization problem for maximum likelihood estimation. The functional form was $F(x) = F_+(x - x_0)$ if $x > x_0$, and $F(x) = -F_-(x - x_0)$ if $x \leq x_0$. The parameters $F_+ = 5.0$, $F_- = 25.0$, and $x_0 = 0.06$ were used, and this formulation appears robust: varying $F_+$ and $F_-$ by two-fold, changing $x_0$ to zero, or using a double exponential fit for $x > x_0$, did not change the resulting corrected data beyond experimental errors (as estimated in the next section). We optimized (3) by performing a grid search of $\zeta = \gamma/(1-\gamma)$ from 0.0 to 2.0 in 0.05 increments and iteratively solving for $\alpha$ and then $\beta$ (through relations obtained by $dL/d\alpha = 0$ and $dL/d\beta = 0$) until convergence. Applying the maximum likelihood values of $\alpha$, $\beta$, and $\gamma$ in (2) gave the corrected SHAPE reactivities. The algorithm is available in the function *overmod_and_background_correct_logL.m* within the freely available HiTRACE software package (*3*).

*Averaging across replicates, estimation of errors, and normalization*

The acquisition of multiple replicates across several experiments permitted high-quality final averaged data $\overline{S}_i$ with error estimates $\overline{\sigma}_i$. To carry out the averaging, we noted that individual experiments might have different levels of measurement precision, and the variance of measurements within each experiment provide an estimate of that precision. These estimates, however, do not include systematic errors that differ between experiments, e.g., differing fluorescent backgrounds in different capillary electrophoresis instruments. We therefore carried out a two-part averaging. First, the data within each experiment $j = 1, 2, .. M$ were averaged to give $S_i^j$ and $\sigma_i^j$. As an example, suppose we have available 20 replicate measurements of each background/overmodification-corrected SHAPE profile $s_i^k$, where $k = 1$ to 20. Suppose these data were measured for 4 independently prepared RNAs across $M = 5$ different days/experiments. Then each $k$ is the member of one and only one of 5 subsets $E_j$. Let $N_j$ be the number of RNAs in set $E_j$ (here $N_j = 4$). For $j = 1, 2, \ldots$ 5,

$$S_i^j = \frac{1}{N_j} \sum_{k \in S_j} s_i^k \tag{4}$$

The estimated errors for these SHAPE data are:

$$\sigma_i^j = \left( \frac{1}{N_j} \sum_{k \in S_j} \left[ s_i^k - S_i^j \right]^2 \right)^{1/2} \tag{5}$$

To combine measurements across multiple experiments, these merged data were averaged, with the inclusion of a position-dependent scale-factor $\alpha_i$ that accounts for additional sources of experiment-to-experiment error. Explicitly, the assumed likelihood model was:

$$L(S_i) = \prod_j \frac{1}{2\pi\alpha_i\sigma_i^j} e^{-\left(S_i - S_i^j\right)^2 / 2\left(\alpha_i\sigma_i^j\right)^2} \tag{6}$$

This gives maximum-likelihood combined signal values $\overline{S}_i$ and final Gaussian errors $\overline{\sigma}_i$ of:

$$\overline{S}_i = \frac{\sum_j \left[ S_i^j / \left(\sigma_i^j\right)^2 \right]}{\sum_j \left[ 1 / \left(\sigma_i^j\right)^2 \right]} \tag{7}$$

$$\overline{\sigma}_i = \alpha_i \left( \sum_j \left[ 1 / \left(\sigma_i^j\right)^2 \right] \right)^{-1/2}.$$

Here, $\alpha_i$ is a scale factor and is again determined by optimizing the likelihood:

$$\alpha_i = \frac{1}{M}\sum_{j=1}^{M}\left[\left(\overline{s_i} - s_i^j\right)^2 / \left(\sigma_i^j\right)^2\right] \tag{8}$$

In practice, to obtain a robust estimate of this error scale factor, the average in (8) is taken across a 5-nucleotide window of bands around each nucleotide *i*. An example of this averaging is given in SI Fig. S3. These data, averaged across multiple replicates, were then normalized following a previously described procedure that was found to be optimal for *E. coli* ribosomal RNA (*1*). Briefly, the data sets were divided by a normalization factor, determined as the average of the top tenth percentile of band intensities. 'Outliers', identified as band intensities that exceeded 1.5 times the interquartile range, were removed before determining this factor. The resulting values lie mostly between 0 and 2 (see e.g., main text Fig. 1). The overall algorithm is available in the function *get_average_standard_state.m* within the freely available HiTRACE software package (*3*).

## Table S1. Benchmark for SHAPE-directed secondary structure modeling.

| RNA, source | Solution conditions[a] | Replicates[b] | Expts[b] | Off-set[c] | PDB[d] | Sequence & Secondary Structure[e] |
|---|---|---|---|---|---|---|
| $tRNA^{phe}$,<br>*E. coli* | Standard | 14 | 7 | -15 | **1L0U**<br>1EHZ | ggaacaaacaaaacaGCGGAUUUAGCUCAGUUGGGAGAGCGCCAGACUGAAGAUCUG<br>GAGGUCCUGUGUUCGAUCCACAGAAUUCGCACCAaaaccaaagaaacaacaacaaca<br>ac<br>...............(((((((..((((........)))).((((.........)))<br>).....(((((.......))))))))))))...........................<br>.. |
| P4-P6 domain,<br>*Tetrahymena* ribozyme | Standard[f] | 28 | 11 | 89 | **1GID**<br>1L8V<br>1HR2<br>2R8S | ggccaaaacaacgGAAUUGCGGGAAAGGGGUCAACAGCCGUUCAGUACCAAGUCUCA<br>GGGGAAACUUUGAGAUGGCCUUGCAAAGGGUAUGGUAAUAAGCUGACGGACAUGGUC<br>CUAACCACGCAGCCAAGUCCUAAGUCAACAGAUCUUCUGUUGAUAUGGAUGCAGUUC<br>Aaaaccaaaccaaagaaacaacaacaacaac<br>.................((((((...((((((.....(((.((((.(((..((((((<br>(((....)))))))))..((.......))....)))......)))))))....))))<br>))..)).))))((...((((...(((((((((...)))))))))..))))...))..<br>............................... |
| 5S rRNA,<br>*E. coli* | Standard | 12 | 6 | -20 | **3OFC**<br>3OAS<br>3ORB<br>2WWQ<br>… | ggaaaggaaagggaaagaaaUGCCUGGCGGCCGUAGCGCGGUGGUCCCACCUGACCC<br>CAUGCCGAACUCAGAAGUGAAACGCCGUAGCGCCGAUGGUAGUGUGGGGUCUCCCCA<br>UGCGAGAGUAGGGAACUGCCAGGCAUaaaacaaaacaaagaaacaacaacaacaac<br>.....................(((((((((.....((((((((....(((((((...<br>..........))))..)))...)))))).)).((.......((((((((...)))))<br>))).......))...)))))))))................................ |
| Adenine riboswitch,<br>*V. vulnificus*<br>(*add*) | Standard +<br>5 mM adenine | 19 | 6 | -8 | **1Y26**<br>1Y27<br>2G9C<br>3GO2<br>… | ggaaaggaaagggaaagaaaCGCUUCAUAUAAUCCUAAUGAUAUGGUUUGGGAGUUU<br>CUACCAAGAGCCUUAAACUCUUGAUUAUGAAGUGaaaacaaaacaaagaaacaacaa<br>caacaac<br>....................(((((((((...((((((.........))))))....<br>....((((((.......))))))..)))))))))......................<br>....... |
| c-di-GMP riboswitch,<br>*V. cholerae*<br>(*VC1722*) | Standard +<br>10 µM cyclic di-guanosine mono-phosphate | 15 | 6 | 0 | **3MXH**<br>3IWN<br>3MUV<br>3MUT<br>… | ggaaaaauGUCACGCACAGGGCAAACCAUUCGAAAGAGUGGGACGCAAAGCCUCCGG<br>CCUAAACCAGAAGACAUGGUAGGUAGCGGGGUUACCGAUGGCAAAAUGcauacaaac<br>caaagaaacaacaacaacaac<br>..........((((......((...((((((....))))))...))...(((.((((<br>((((..((.........)))))))..))))))...)).)).................<br>..................... |
| Glycine riboswitch,<br>*F. nucleatum* | Standard +<br>10 mM glycine | 22 | 8 | -10 | **3P49** | ggacagagagGAUAUGAGGAGAGAUUUCAUUUUAAUGAAACACCGAAGAAGUAAAUC<br>UUUCAGGUAAAAAGGACUCAUAUUGGACGAACCUCUGGAGAGCUUAUCUAAGAGAUA<br>ACACCGAAGGAGCAAAGCUAAUUUUAGCCUAAACUCUCAGGUAAAAGGACGGAGaaa<br>acacaacaaagaaacaacaacaacaac<br>..........((((((((......((((((....)))))).(((....(((.....)<br>))...)))........))))))))........(((((......(((((.....))))<br>).(((...((((.....(((....)))......))))..))).......)))))...<br>........................... |

[a] Standard conditions are: 10 mM $MgCl_2$, 50 mM Na-HEPES, pH 8.0 at 24 °C.

[b] All data average over experiments carried out on at least four different days to minimize systematic errors in sample preparations; within each experiment, two or more independently prepared and purified RNA stocks were assayed.

[c] Number added to sequence index to yield numbering used in previous biophysical studies, and in Figs. 1 and 2 of the main text.

[d] The first listed PDB ID was the source of the assumed crystallographic secondary structure; other listed IDs contain sequence variants, different complexes, or different crystallographic space groups and confirm this structure.

[e] In the sequence, lowercase symbols denote 5´and 3´ buffer sequences, including primer binding site (last 20 nucleotides). In all cases, designs were checked in RNAstructure and ViennaRNA to give negligible base pairing between added sequences and target domain. Structure is given in dot-bracket notation, and here denotes Watson/Crick base pairs for which there is crystallographic evidence. Only helices with two or more base pairs are included. For the adenine riboswitch, a two-base-pair helix [25-50, 26-49] that is not nested in the given secondary structure and involved in an extensive non-canonical loop-loop interaction is not included.

[f] Additional measurements were carried out with 30% methylpentanediol (MPD) due to reports that its presence in crystallization buffer can change SHAPE reactivity of the P4-P6 RNA (*4*). Measurements with MPD (10 replicates) gave different reactivities in the P5abc region; final SHAPE-directed secondary structure models mispredicted an additional helix compared to models guided by no-MPD data.

**Table S2. Base-pair-level statistics of secondary structure recovery by *RNAstructure* with and without SHAPE data.** TP=true positives; FP=false positives. TP´and FP´ are the same, but allowing matches of base pair (i,j) with (i±1, j±1).

| RNA | Len. | Number of base pairs | | | | | | | | |
|---|---|---|---|---|---|---|---|---|---|---|
| | | Cryst[a] | RNAstructure | | | | + SHAPE | | | |
| | | | TP | FP | TP´ | FP´ | TP | FP | TP´ | FP´ |
| tRNA[phe] | 76 | 20 | 12 | 12 | 12 | 12 | 15 | 6 | 15 | 4 |
| P4-P6 RNA | 158 | 48 | 44 | 9 | 48 | 2 | 44 | 7 | 46 | 2 |
| 5S rRNA | 118 | 34 | 9 | 31 | 9 | 31 | 32 | 7 | 32 | 7 |
| Adenine ribosw. | 71 | 21 | 15 | 10 | 15 | 10 | 21 | 2 | 21 | 2 |
| c-di-GMP ribosw. | 80 | 25 | 21 | 5 | 21 | 3 | 21 | 6 | 21 | 4 |
| Glycine riboswitch | 158 | 40 | 23 | 18 | 23 | 16 | 37 | 7 | 37 | 5 |
| Total | 661 | 188 | 124 | 85 | 128 | 74 | 170 | 35 | 172 | 24 |
| **False negative rate**[b] | | | 34.0% | | 31.9% | | 9.6% | | 8.5% | |
| **False discovery rate**[c] | | | 40.7% | | 36.6% | | 17.1% | | 12.2% | |
| **Sensitivity**[d] | | | 66.0% | | 68.1% | | 90.4% | | 91.5% | |
| **Positive predictive value**[e] | | | 59.3% | | 63.4% | | 82.9% | | 87.8% | |

[a] Cryst = number of helices in crystallographic model.
[b] False negative rate = 1 - TP/Cryst.
[c] False discovery rate = FP/(TP+FP).
[d] Sensitivity = (1 - false negative rate) = TP/Cryst.
[e] Positive predictive value = (1 - false discovery rate) = TP/(TP+FP).

**Table S3. Sources of poor discrimination of correct from incorrect secondary structures**. Thermodynamic energies of base pairs and SHAPE pseudoenergies in kcal/mol, calculated in *RNAstructure*.

| | SHAPE-directed model | | | Crystallographic model[a] | | | Difference[b] | | |
|---|---|---|---|---|---|---|---|---|---|
| RNA | $E_{total}$ | $E_{thermo}$ | $E_{SHAPE}$ | $E_{total}$ | $E_{thermo}$ | $E_{SHAPE}$ | $E_{total}$ | $E_{thermo}$ | $E_{SHAPE}$ |
| tRNA[phe] | -40.1 | -20.3 | -19.8 | -39.6 | -20.5 | -19.1 | -0.5 | 0.2 | -0.7 |
| P4-P6 | -125.6 | -54.8 | -70.8 | -114.9 | -46.4 | -68.5 | -10.7 | -8.4 | -2.3 |
| 5S rRNA | -95.5 | -47.5 | -48.0 | -91.9 | -45.7 | -46.2 | -3.6 | -1.8 | -1.8 |
| Ade ribosw. | -48.2 | -16.6 | -31.6 | -48.2 | -16.6 | -31.6 | 0.0 | 0.0 | 0.0 |
| c-di-GMP ribosw. | -63.6 | -26.3 | -37.3 | -62.7 | -26.4 | -36.3 | -0.9 | 0.1 | -1.0 |
| Gly. ribosw. | -98.5 | -24.8 | -73.7 | -96 | -24.5 | -71.5 | -2.5 | -0.3 | -2.2 |
| *Average* | -78.5 | -31.7 | -46.9 | -75.5 | -30.0 | -45.5 | -3.0 | -1.7 | -1.3 |

[a] For a fair comparison to the SHAPE-directed model, this is the lowest energy secondary structure produced by *RNAstructure* with the same SHAPE data, but forced to contain the crystallographically observed base pairs. For the adenine riboswitch, an 'extra' two-base-pair helix appears in this structure.
[b]Negative values indicate inaccuracy in structure discrimination.
[c] $E_{total}$ and $E_{thermo}$ are derived from from *efn2* (the *RNAstructure* package) run with and without SHAPE data, respectively. $E_{SHAPE}$ is the difference of the two values.

**Table S4. Helix-by-helix bootstrap confidence estimates for the SHAPE-directed model of the HIV-1 RNA genome.** Models were generated by applying RNAstructure 5.3 to SHAPE data from ref. (*5*). Following prior work, the temperature was set to the default (37 °C); slope *m* and intercept *b* of SHAPE pseudoenergy relation were set to 3.0 kcal/mol and -0.6 kcal/mol, respectively; maximum sequence distance between base pairs was set to 600; modeling was carried out for separate subsegments 1-2844, 2836-5722, 5676-6832, 6807-7791, and 7779-9173; and positions at termini of these subsegments (2836-2845, 5676-5724, 6799-6838, 7779-7791, 9171-9173) and in pseudoknotted regions (179-216, bound to tRNALys primer; 255-263 in the dimerization loop DIS; and 74-86 and 408-375, forming the 5´ polyA signal) were forced to remain unpaired. "BP1", "BP2", and "len" give two residues marking the starting base pair of each stem and stem length; "P(boot)" and "BPP" are bootstrap confidence value and maximum Boltzmann probability in the stem as percentages; and "Modeled" gives whether the stem was in the working model of (*5*) and recovered with RNAstructure 5.3 (Y) or not (N), or whether the stem is newly predicted herein (X). Continued on next two pages.

| BP1 | BP2 | len | P(boot) | BPP | Modeled | BP1 | BP2 | len | P(boot) | BPP | Modeled | BP1 | BP2 | len | P(boot) | BPP | Modeled |
|---|---|---|---|---|---|---|---|---|---|---|---|---|---|---|---|---|---|
| 1 | 57 | 3 | 95.5 | 31.1 | Y[a] | 1214 | 1247 | 5 | 89 | 99.8 | Y | 2619 | 2666 | 4 | 96.5 | 100 | Y |
| 5 | 54 | 11 | 100 | 100 | Y[a] | 1223 | 1239 | 6 | 94 | 100 | Y | 2625 | 2661 | 4 | 91.5 | 98.9 | Y |
| 17 | 43 | 5 | 100 | 100 | Y[a] | 1249 | 1263 | 4 | 88.5 | 99.9 | Y | 2629 | 2654 | 10 | 100 | 100 | Y |
| 25 | 38 | 4 | 100 | 100 | Y[a] | 1350 | 1727 | 2 | 17 | 2.4 | Y | 2667 | 2686 | 3 | 47 | 6.5 | Y |
| 58 | 104 | 8 | 99.5 | 100 | Y | 1353 | 1724 | 5 | 20.5 | 2.8 | Y | 2671 | 2683 | 3 | 41 | 5.8 | Y |
| 67 | 94 | 3 | 99.5 | 100 | Y | 1360 | 1394 | 4 | 19.5 | 2 | Y | 2727 | 2745 | 4 | 89 | 100 | Y |
| 70 | 90 | 4 | 99.5 | 100 | Y | 1375 | 1387 | 5 | 69.5 | 95.3 | Y | 2731 | 2740 | 2 | 76 | 95.2 | Y |
| 106 | 343 | 9 | 40 | 98.5 | Y | 1396 | 1558 | 4 | 37 | 5.5 | Y | 2781 | 2802 | 7 | 92 | 100 | Y |
| 125 | 223 | 7 | 26.5 | 95.4 | Y | 1401 | 1554 | 3 | 38 | 5.6 | Y | 2811 | 2835 | 2 | 67 | 98.8 | Y |
| 134 | 178 | 8 | 77.5 | 99.4 | Y | 1405 | 1414 | 2 | 24 | 24.5 | Y | 2814 | 2833 | 6 | 80.5 | 100 | Y |
| 143 | 167 | 2 | 51.5 | 93.2 | Y | 1418 | 1457 | 3 | 29.5 | 21.8 | Y | 2846 | 3381 | 6 | 68.5 | 65.8 | Y |
| 146 | 164 | 2 | 64.5 | 94 | Y | 1421 | 1443 | 7 | 92.5 | 99.7 | Y | 2852 | 3374 | 5 | 53 | 64.6 | Y |
| 148 | 160 | 3 | 79 | 99.7 | Y | 1459 | 1522 | 6 | 54 | 29.4 | Y | 2876 | 3273 | 5 | 10.5 | 22.1 | Y |
| 228 | 334 | 6 | 11 | 27.1 | Y | 1465 | 1515 | 3 | 37.5 | 26.4 | Y | 2892 | 3176 | 5 | 16 | 22.4 | Y |
| 236 | 282 | 3 | 70 | 69.3 | Y | 1469 | 1511 | 3 | 34 | 26.2 | Y | 2908 | 3160 | 4 | 12.5 | 21.2 | Y |
| 243 | 277 | 4 | 93 | 77.2 | Y | 1477 | 1504 | 3 | 18.5 | 20.9 | Y | 2925 | 3129 | 3 | 7.5 | 17.5 | Y |
| 248 | 270 | 7 | 100 | 100 | Y | 1481 | 1500 | 4 | 14.5 | 15.6 | Y | 2939 | 3088 | 5 | 2.5 | 7.9 | Y |
| 283 | 299 | 3 | 37 | 70 | Y | 1531 | 1541 | 3 | 69.5 | 90.7 | Y | 2946 | 2953 | 2 | 22.5 | 18.1 | Y |
| 286 | 295 | 3 | 34.5 | 69.3 | Y | 1568 | 1707 | 10 | 100 | 100 | Y[B] | 2972 | 3038 | 5 | 8 | 8.5 | Y |
| 312 | 325 | 5 | 98 | 100 | Y | 1583 | 1694 | 3 | 53.5 | 68.3 | Y[B] | 2995 | 3006 | 4 | 38.5 | 46.6 | Y |
| 363 | 750 | 5 | 4.5 | 1.6 | Y | 1590 | 1683 | 6 | 48.5 | 77.8 | Y[B] | 3015 | 3023 | 3 | 28 | 45.9 | Y |
| 399 | 484 | 9 | 98 | 100 | Y | 1598 | 1640 | 2 | 35.5 | 65.8 | Y[B] | 3040 | 3058 | 3 | 36 | 56.1 | Y |
| 501 | 526 | 6 | 33 | 87 | Y | 1604 | 1636 | 8 | 78 | 98.7 | Y[B] | 3044 | 3055 | 2 | 35 | 55.9 | Y |
| 510 | 518 | 2 | 22 | 62 | Y | 1615 | 1625 | 3 | 97 | 100 | Y[B] | 3089 | 3105 | 2 | 24.5 | 17.7 | Y |
| 582 | 657 | 2 | 41.5 | 68.2 | Y | 1645 | 1672 | 12 | 100 | 100 | Y[B] | 3093 | 3101 | 3 | 41 | 38 | Y |
| 586 | 652 | 8 | 66.5 | 99.7 | Y | 1760 | 1785 | 4 | 27.5 | 28.4 | Y | 3140 | 3149 | 3 | 60.5 | 81 | Y |
| 595 | 625 | 4 | 58.5 | 91.6 | Y | 1767 | 1779 | 2 | 30 | 27.2 | Y | 3178 | 3190 | 3 | 40.5 | 91 | Y |
| 599 | 616 | 5 | 93 | 99.8 | Y | 1813 | 1916 | 6 | 53 | 94.9 | Y | 3205 | 3223 | 3 | 93 | 99.4 | Y |
| 628 | 636 | 3 | 36.5 | 83.8 | Y | 1823 | 1849 | 6 | 66 | 90.9 | Y | 3243 | 3256 | 4 | 50.5 | 47.1 | Y |
| 678 | 741 | 4 | 4 | 4.4 | Y | 1829 | 1842 | 3 | 88.5 | 99.4 | Y | 3285 | 3334 | 1 | 7 | 17.7 | Y |
| 683 | 691 | 2 | 25.5 | 12.1 | Y | 1862 | 1881 | 5 | 78 | 99.9 | Y | 3287 | 3332 | 6 | 15.5 | 22.7 | Y |
| 693 | 722 | 6 | 73.5 | 99.6 | Y | 1991 | 2326 | 9 | 57.5 | 100 | Y | 3297 | 3322 | 5 | 34 | 25.9 | Y |
| 702 | 714 | 2 | 59 | 93.4 | Y | 2015 | 2121 | 8 | 99.5 | 100 | Y | 3336 | 3358 | 5 | 8.5 | 25.4 | Y |
| 752 | 1172 | 5 | 22 | 36.9 | Y | 2024 | 2112 | 10 | 99 | 100 | Y | 3344 | 3351 | 2 | 9.5 | 22.3 | Y |
| 795 | 849 | 9 | 87.5 | 99.9 | Y | 2042 | 2070 | 7 | 82 | 99.9 | Y | 3393 | 3402 | 2 | 19 | 11.7 | Y |
| 821 | 831 | 4 | 88 | 98.2 | Y | 2051 | 2062 | 3 | 35 | 56.3 | Y | 3404 | 3943 | 4 | 11.5 | 5 | Y |
| 871 | 913 | 2 | 56 | 84.1 | Y | 2072 | 2082 | 3 | 25.5 | 23.3 | Y | 3410 | 3936 | 4 | 15.5 | 32.1 | Y |
| 874 | 911 | 6 | 77 | 99.3 | Y | 2135 | 2144 | 3 | 35.5 | 82.6 | Y | 3414 | 3931 | 7 | 18.5 | 33.8 | Y |
| 921 | 964 | 5 | 51 | 99.1 | Y | 2145 | 2171 | 6 | 58 | 91.6 | Y | 3423 | 3924 | 4 | 12 | 31.7 | Y |
| 926 | 935 | 3 | 20.5 | 68.1 | Y | 2201 | 2238 | 7 | 38.5 | 89.8 | Y | 3427 | 3496 | 4 | 16 | 37 | Y |
| 946 | 957 | 2 | 30 | 68.5 | Y | 2245 | 2260 | 4 | 35 | 91.5 | Y | 3432 | 3439 | 2 | 35 | 53.8 | Y |
| 1028 | 1064 | 5 | 67 | 82.2 | Y | 2268 | 2310 | 5 | 88 | 100 | Y | 3458 | 3468 | 3 | 44.5 | 89.4 | Y |
| 1076 | 1100 | 3 | 99 | 99.9 | Y | 2273 | 2301 | 5 | 88 | 99.8 | Y | 3497 | 3917 | 5 | 15 | 39.3 | Y |
| 1080 | 1097 | 5 | 100 | 100 | Y | 2279 | 2296 | 1 | 88 | 49.4 | Y | 3523 | 3653 | 6 | 15.5 | 5.6 | Y |
| 1102 | 1142 | 5 | 85 | 99.6 | Y | 2378 | 2429 | 8 | 68 | 99.2 | Y | 3532 | 3644 | 2 | 8 | 4 | Y |
| 1110 | 1137 | 3 | 84 | 99.9 | Y | 2391 | 2421 | 3 | 32 | 69.7 | Y | 3535 | 3550 | 3 | 31 | 53.5 | Y |
| 1116 | 1132 | 6 | 75 | 98.8 | Y | 2547 | 2778 | 9 | 52 | 0 | Y | 3581 | 3593 | 4 | 11 | 3.8 | Y |
| 1177 | 1312 | 5 | 64.5 | 39.2 | Y | 2558 | 2576 | 3 | 16 | 1.7 | Y | 3606 | 3639 | 2 | 19 | 41.4 | Y |
| 1183 | 1306 | 5 | 68.5 | 39.2 | Y | 2596 | 2712 | 5 | 31.5 | 0.4 | Y | 3609 | 3636 | 6 | 25 | 50.7 | Y |
| 1193 | 1299 | 6 | 83 | 99.2 | Y | 2603 | 2705 | 4 | 30.5 | 0.4 | Y | 3692 | 3907 | 6 | 18 | 61.6 | Y |
| 1214 | 1247 | 5 | 89 | 99.8 | Y | 2619 | 2666 | 4 | 96.5 | 100 | Y | 3699 | 3800 | 4 | 35 | 84.9 | Y |
| 1223 | 1239 | 6 | 94 | 100 | Y | 2625 | 2661 | 4 | 91.5 | 98.9 | Y | 3704 | 3759 | 7 | 94 | 99.9 | Y |
| 1249 | 1263 | 4 | 88.5 | 99.9 | Y | 2629 | 2654 | 10 | 100 | 100 | Y | 3818 | 3899 | 9 | 89.5 | 100 | Y |
| 1350 | 1727 | 2 | 17 | 2.4 | Y | 2667 | 2686 | 3 | 47 | 6.5 | Y | 3840 | 3890 | 6 | 91 | 100 | Y |
| 1353 | 1724 | 5 | 20.5 | 2.8 | Y | 2671 | 2683 | 3 | 41 | 5.8 | Y | 3857 | 3872 | 4 | 82.5 | 99.8 | Y |

[a] 5´ TAR element.

[b] gag-pol region.

| BP1 | BP2 | len | P(boot) | BPP | Modeled | BP1 | BP2 | len | P(boot) | BPP | Modeled | BP1 | BP2 | len | P(boot) | BPP | Modeled |
|---|---|---|---|---|---|---|---|---|---|---|---|---|---|---|---|---|---|
| 4204 | 4349 | 8 | 96.5 | 100 | Y | 5499 | 5519 | 3 | 47 | 79.8 | Y | 6538 | 6595 | 4 | 66 | 95 | Y |
| 4280 | 4299 | 4 | 91.5 | 96.8 | Y | 5503 | 5515 | 4 | 40 | 75.7 | Y | 6543 | 6590 | 3 | 45 | 33.4 | Y |
| 4301 | 4316 | 4 | 32.5 | 28.6 | Y | 5530 | 5581 | 5 | 24.5 | 69.1 | Y | 6546 | 6586 | 6 | 96 | 100 | Y |
| 4362 | 4503 | 7 | 45.5 | 52.4 | Y | 5536 | 5576 | 3 | 27 | 69.5 | Y | 6552 | 6579 | 4 | 90.5 | 99.7 | Y |
| 4447 | 4488 | 4 | 44 | 40.8 | Y | 5542 | 5571 | 2 | 31.5 | 70.5 | Y | 6618 | 6635 | 4 | 21.5 | 59.7 | Y |
| 4456 | 4479 | 2 | 34 | 31.5 | Y | 5545 | 5568 | 3 | 36 | 72.5 | Y | 6623 | 6630 | 2 | 18.5 | 56.4 | Y |
| 4458 | 4473 | 3 | 56 | 55.8 | Y | 5549 | 5565 | 2 | 15 | 49.1 | Y | 6706 | 6762 | 2 | 26 | 43.1 | Y |
| 4490 | 4496 | 2 | 32 | 36 | Y | 5600 | 5644 | 2 | 59.5 | 61.4 | Y | 6711 | 6757 | 9 | 90 | 100 | Y |
| 4551 | 5036 | 10 | 81 | 100 | Y | 5605 | 5639 | 6 | 95 | 100 | Y | 6839 | 7188 | 4 | 52 | 79.7 | Y |
| 4573 | 4586 | 4 | 26.5 | 53.3 | Y | 5612 | 5632 | 1 | 95 | 76.9 | Y | 6846 | 7179 | 2 | 71 | 97.1 | Y |
| 4588 | 4934 | 9 | 71 | 99.9 | Y | 5614 | 5631 | 4 | 95 | 99.9 | Y | 6850 | 7176 | 5 | 88 | 99.7 | Y |
| 4601 | 4914 | 8 | 87 | 100 | Y | 5725 | 6314 | 3 | 27.5 | 25.2 | Y | 6864 | 7113 | 5 | 10.5 | 16.8 | Y |
| 4614 | 4902 | 2 | 16 | 24.3 | Y | 5745 | 6243 | 10 | 97 | 99.9 | Y | 6870 | 6886 | 6 | 26 | 30.7 | Y |
| 4642 | 4674 | 3 | 28 | 85.4 | Y | 5763 | 6147 | 4 | 15.5 | 12.1 | Y | 6893 | 7077 | 1 | 19 | 7.9 | Y |
| 4646 | 4670 | 4 | 35 | 94.4 | Y | 5770 | 6142 | 6 | 37.5 | 16.9 | Y | 6894 | 7075 | 1 | 20 | 7.2 | Y |
| 4694 | 4732 | 3 | 31.5 | 86.2 | Y | 5793 | 6013 | 7 | 96.5 | 100 | Y | 6895 | 7073 | 7 | 97.5 | 100 | Y |
| 4698 | 4728 | 4 | 31.5 | 86.4 | Y | 5803 | 6004 | 5 | 91.5 | 99.9 | Y | 6904 | 7056 | 10 | 98 | 100 | Y |
| 4754 | 4770 | 6 | 100 | 100 | Y | 5816 | 5830 | 6 | 98.5 | 100 | Y | 6923 | 7040 | 4 | 78 | 97.9 | Y |
| 4797 | 4899 | 7 | 97.5 | 100 | Y | 5846 | 5861 | 6 | 98.5 | 100 | Y[c] | 6938 | 6956 | 5 | 81.5 | 96.8 | Y |
| 4807 | 4822 | 3 | 99.5 | 100 | Y | 5867 | 5997 | 7 | 94 | 100 | Y | 6964 | 6973 | 3 | 33.5 | 66.3 | Y |
| 4829 | 4891 | 9 | 100 | 100 | Y | 5874 | 5989 | 5 | 93.5 | 99.9 | Y | 6983 | 7016 | 3 | 13.5 | 33.9 | Y |
| 4840 | 4856 | 6 | 100 | 100 | Y | 5887 | 5894 | 2 | 26.5 | 63.6 | Y | 6988 | 7009 | 2 | 34.5 | 70.7 | Y |
| 4938 | 4999 | 6 | 81 | 100 | Y | 5896 | 5980 | 6 | 57.5 | 89.9 | Y | 6991 | 7007 | 4 | 40 | 80.5 | Y |
| 4951 | 4985 | 5 | 75.5 | 99.9 | Y | 5904 | 5913 | 2 | 36.5 | 63.6 | Y | 7079 | 7099 | 7 | 94 | 100 | Y |
| 4960 | 4980 | 6 | 99.5 | 100 | Y | 5915 | 5973 | 3 | 21 | 46.9 | Y | 7114 | 7136 | 7 | 94 | 98.9 | Y |
| 5010 | 5022 | 4 | 39.5 | 92.6 | Y | 5927 | 5962 | 4 | 24 | 41.4 | Y | 7150 | 7170 | 5 | 13 | 16.8 | Y |
| 5070 | 5100 | 6 | 19 | 47.2 | Y | 5931 | 5952 | 4 | 69 | 99.5 | Y | 7245 | 7599 | 1 | 66.5 | 80 | Y |
| 5083 | 5094 | 4 | 51 | 87.8 | Y | 5936 | 5947 | 4 | 78.5 | 100 | Y | 7247 | 7597 | 5 | 84 | 99.9 | Y |
| 5114 | 5132 | 4 | 73.5 | 4.8 | Y | 6024 | 6135 | 3 | 19.5 | 4.2 | Y | 7256 | 7590 | 8 | 99 | 100 | Y |
| 5139 | 5675 | 2 | 10 | 0.9 | Y | 6048 | 6066 | 5 | 67.5 | 92.4 | Y | 7272 | 7578 | 11 | 99.5 | 100 | Y |
| 5143 | 5673 | 7 | 20.5 | 1.7 | Y | 6072 | 6125 | 3 | 13 | 2.4 | Y | 7283 | 7566 | 3 | 99 | 100 | Y |
| 5154 | 5204 | 4 | 43 | 57.5 | Y | 6076 | 6121 | 2 | 12.5 | 2.4 | Y | 7291 | 7557 | 7 | 98.5 | 100 | Y |
| 5166 | 5194 | 5 | 75.5 | 85.2 | Y | 6078 | 6118 | 2 | 15.5 | 2.5 | Y | 7305 | 7538 | 7 | 94.5 | 100 | Y |
| 5206 | 5396 | 2 | 29 | 40.6 | Y | 6083 | 6113 | 3 | 22 | 4.5 | Y | 7312 | 7530 | 3 | 94 | 100 | Y |
| 5209 | 5394 | 5 | 77 | 100 | Y | 6092 | 6102 | 3 | 6.5 | 3.6 | Y | 7316 | 7526 | 5 | 94 | 100 | Y |
| 5216 | 5384 | 8 | 82 | 100 | Y | 6149 | 6159 | 2 | 2 | 5.4 | Y | 7321 | 7520 | 4 | 99.5 | 99.9 | Y |
| 5234 | 5265 | 3 | 29 | 56.9 | Y | 6185 | 6200 | 2 | 40.5 | 15.5 | Y | 7325 | 7515 | 4 | 99.5 | 100 | Y |
| 5239 | 5261 | 5 | 69 | 98.7 | Y | 6270 | 6290 | 6 | 69 | 98 | Y | 7333 | 7508 | 4 | 100 | 100 | Y |
| 5267 | 5297 | 6 | 65.5 | 97.9 | Y | 6328 | 6798 | 3 | 35.5 | 30.6 | Y | 7337 | 7503 | 5 | 100 | 100 | Y |
| 5273 | 5283 | 4 | 52 | 82.1 | Y | 6332 | 6375 | 6 | 20 | 34.8 | Y | 7343 | 7408 | 6 | 92.5 | 100 | Y |
| 5303 | 5343 | 4 | 50.5 | 96.6 | Y | 6381 | 6780 | 4 | 16 | 46.1 | Y | 7350 | 7378 | 1 | 67 | 95.5 | Y |
| 5307 | 5337 | 4 | 40 | 94.7 | Y | 6385 | 6775 | 3 | 15 | 47.7 | Y | 7353 | 7375 | 3 | 77 | 100 | Y |
| 5311 | 5331 | 6 | 76.5 | 98.1 | Y | 6393 | 6767 | 2 | 13 | 44.3 | Y | 7356 | 7371 | 2 | 82 | 100 | Y |
| 5349 | 5370 | 2 | 17 | 32.1 | Y | 6398 | 6704 | 5 | 30.5 | 88.6 | Y | 7358 | 7368 | 4 | 94.5 | 100 | Y |
| 5352 | 5367 | 1 | 17 | 20.1 | Y | 6416 | 6695 | 4 | 25 | 86.3 | Y | 7383 | 7399 | 6 | 100 | 100 | Y |
| 5355 | 5364 | 3 | 26 | 43 | Y | 6421 | 6521 | 8 | 26.5 | 85.8 | Y | 7411 | 7428 | 5 | 78 | 99.5 | Y |
| 5404 | 5419 | 5 | 99 | 100 | Y | 6432 | 6513 | 6 | 28 | 84.5 | Y | 7438 | 7464 | 3 | 63.5 | 98.1 | Y |
| 5425 | 5438 | 3 | 32.5 | 52.5 | Y | 6456 | 6467 | 4 | 41 | 88.4 | Y | 7443 | 7459 | 7 | 78 | 99.5 | Y |
| 5440 | 5450 | 3 | 28.5 | 40.4 | Y | 6475 | 6497 | 5 | 18 | 36.7 | Y | 7468 | 7493 | 9 | 100 | 100 | Y |
| 5473 | 5650 | 5 | 1 | 2.3 | Y | 6536 | 6598 | 2 | 39.5 | 67.5 | Y | 7601 | 7616 | 4 | 29 | 58.2 | Y |
| 5499 | 5519 | 3 | 47 | 79.8 | Y | 6538 | 6595 | 4 | 66 | 95 | Y | 7627 | 7636 | 2 | 48.5 | 58.4 | Y |
| 5503 | 5515 | 4 | 40 | 75.7 | Y | 6543 | 6590 | 3 | 45 | 33.4 | Y | 7647 | 7692 | 6 | 59.5 | 81.6 | Y |
| 5530 | 5581 | 5 | 24.5 | 69.1 | Y | 6546 | 6586 | 6 | 96 | 100 | Y | 7705 | 7770 | 6 | 67.5 | 99.8 | Y |
| 5536 | 5576 | 3 | 27 | 69.5 | Y | 6552 | 6579 | 4 | 90.5 | 99.7 | Y | 7712 | 7764 | 2 | 58 | 73.9 | Y |
| 5542 | 5571 | 2 | 31.5 | 70.5 | Y | 6618 | 6635 | 4 | 21.5 | 59.7 | Y | 7716 | 7760 | 2 | 61 | 96.4 | Y |

[c]signal-peptide stem at 5′end of gp120.

| BP1 | BP2 | len | P(boot) | BPP | Modeled | BP1 | BP2 | len | P(boot) | BPP | Modeled | BP1 | BP2 | len | P(boot) | BPP | Modeled |
|---|---|---|---|---|---|---|---|---|---|---|---|---|---|---|---|---|---|
| 7939 | 8107 | 2 | 13.5 | 76.6 | Y | 1341 | 1795 | 2 | 1.5 | 0.1 | N | 383 | 393 | 2 | 18.5 | 25.9 | X |
| 7941 | 8051 | 2 | 12 | 24.3 | Y | 1346 | 1790 | 4 | 1 | 0 | N | 380 | 396 | 2 | 14.5 | 17.6 | X |
| 7944 | 8049 | 5 | 50 | 97.8 | Y | 1729 | 1757 | 6 | 15 | 18.3 | N | 533 | 549 | 4 | 31 | 38.2 | X |
| 7949 | 8043 | 3 | 45 | 97.3 | Y | 1737 | 1749 | 3 | 26.5 | 21.4 | N | 972 | 979 | 2 | 17.5 | 4.6 | X |
| 7954 | 8039 | 3 | 19.5 | 31.9 | Y | 1796 | 1946 | 7 | 23.5 | 0.8 | N | 968 | 983 | 3 | 17.5 | 4.4 | X |
| 7969 | 8024 | 4 | 39 | 89.8 | Y | 1809 | 1920 | 2 | 20.5 | 2 | N | 966 | 984 | 2 | 8 | 1.7 | X |
| 7991 | 8015 | 5 | 94.5 | 100 | Y | 1811 | 1917 | 1 | 27.5 | 2.1 | N | 993 | 999 | 2 | 5 | 1.3 | X |
| 7997 | 8009 | 3 | 72 | 98.3 | Y | 1922 | 1930 | 3 | 29.5 | 18.2 | N | 987 | 1008 | 5 | 9 | 1.7 | X |
| 8053 | 8077 | 5 | 43 | 66.5 | Y | 1948 | 2545 | 7 | 7 | 0 | N | 770 | 1014 | 4 | 14 | 3.1 | X |
| 8059 | 8071 | 3 | 65.5 | 98.6 | Y | 2328 | 2348 | 3 | 17 | 1.7 | N | 765 | 1019 | 4 | 13.5 | 3 | X |
| 8083 | 8099 | 3 | 7.5 | 26.2 | Y | 2333 | 2343 | 3 | 17.5 | 1.6 | N | 761 | 1027 | 3 | 14.5 | 3.1 | X |
| 8169 | 8205 | 4 | 33.5 | 80.1 | Y | 2352 | 2520 | 8 | 7 | 0.5 | N | 1731 | 1752 | 5 | 40.5 | 74 | X |
| 8173 | 8194 | 3 | 38.5 | 64.4 | Y | 2363 | 2376 | 4 | 20.5 | 2.3 | N | 1795 | 1920 | 5 | 3.5 | 0.4 | X |
| 8309 | 8326 | 2 | 69 | 8.3 | Y | 2432 | 2484 | 6 | 30.5 | 1.1 | N | 1787 | 1925 | 3 | 3 | 0.3 | X |
| 8648 | 8667 | 3 | 45.5 | 66.1 | Y | 2448 | 2472 | 4 | 29.5 | 1.1 | N | 1345 | 1930 | 3 | 8.5 | 0 | X |
| 8651 | 8663 | 4 | 67.5 | 99.8 | Y | 2486 | 2497 | 2 | 2.5 | 0.2 | N | 2350 | 2357 | 2 | 19.5 | 35.8 | X |
| 8669 | 8679 | 3 | 98 | 100 | Y | 2714 | 2725 | 4 | 20 | 20.2 | N | 2348 | 2360 | 2 | 15 | 25.1 | X |
| 8686 | 9009 | 7 | 19 | 5.9 | Y | 3945 | 4518 | 4 | 7 | 0 | N | 2343 | 2366 | 5 | 17 | 47.2 | X |
| 8694 | 8999 | 4 | 12 | 5.7 | Y | 3958 | 3970 | 4 | 12.5 | 33.2 | N | 2430 | 2436 | 2 | 38 | 72.8 | X |
| 8700 | 8803 | 6 | 44.5 | 99 | Y | 4057 | 4068 | 2 | 31 | 24.5 | N | 1941 | 2453 | 13 | 51 | 0.1 | X |
| 8723 | 8751 | 11 | 96 | 100 | Y | 4539 | 5135 | 3 | 7.5 | 0 | N | 2496 | 2513 | 2 | 67.5 | 90.6 | X |
| 8753 | 8773 | 8 | 98.5 | 100 | Y | 7540 | 7546 | 2 | 22.5 | 43.4 | N | 2492 | 2517 | 3 | 65.5 | 92.3 | X |
| 8779 | 8790 | 3 | 34.5 | 81.8 | Y | 7638 | 7778 | 4 | 6.5 | 1.7 | N | 2477 | 2531 | 4 | 20 | 13.2 | X |
| 8807 | 8994 | 6 | 13 | 5.8 | Y | 7644 | 7774 | 3 | 3.5 | 0.6 | N | 2469 | 2544 | 8 | 37 | 17 | X |
| 8817 | 8985 | 5 | 31 | 6 | Y | 8226 | 8268 | 5 | 1 | 0 | N | 2721 | 2750 | 4 | 35 | 43 | X |
| 8830 | 8973 | 4 | 35 | 5.8 | Y | 8275 | 8348 | 5 | 26.5 | 3.8 | N | 2899 | 3167 | 3 | 5.5 | 9.8 | X |
| 8838 | 8965 | 2 | 21 | 4.7 | Y | 8282 | 8341 | 5 | 51 | 4.5 | N | 3765 | 3773 | 2 | 16.5 | 26.8 | X |
| 8851 | 8859 | 2 | 25 | 10.1 | Y | 8290 | 8308 | 3 | 19 | 1.7 | N | 3961 | 3970 | 2 | 5.5 | 19.6 | X |
| 8867 | 8906 | 10 | 99.5 | 100 | Y | 8358 | 8684 | 5 | 0.5 | 0.1 | N | 3958 | 4047 | 2 | 5.5 | 9 | X |
| 8878 | 8896 | 4 | 99 | 100 | Y | 8371 | 8412 | 1 | 5 | 1.1 | N | 3945 | 4058 | 4 | 11.5 | 30.8 | X |
| 8884 | 8890 | 1 | 98 | 91.6 | Y | 8373 | 8411 | 6 | 12.5 | 6.3 | N | 4131 | 4518 | 2 | 22 | 0.4 | X |
| 8912 | 8927 | 5 | 69.5 | 87.2 | Y | 8379 | 8404 | 1 | 11.5 | 4.1 | N | 6251 | 6258 | 2 | 24 | 65.6 | X |
| 9042 | 9057 | 5 | 84.5 | 99.7 | Y | 8382 | 8401 | 2 | 10.5 | 4.5 | N | 6390 | 6770 | 2 | 12.5 | 21.2 | X |
| 9059 | 9067 | 2 | 77 | 96.2 | Y | 8417 | 8641 | 7 | 3 | 77.2 | N | 7156 | 7164 | 2 | 13 | 13.4 | X |
| 9074 | 9134 | 5 | 98 | 99.9 | Y | 8432 | 8528 | 4 | 8 | 65.8 | N | 7621 | 7643 | 3 | 19 | 37.9 | X |
| 9080 | 9129 | 11 | 100 | 100 | Y | 8440 | 8519 | 3 | 9.5 | 58.9 | N | 7236 | 7777 | 4 | 47.5 | 95.9 | X |
| 9092 | 9118 | 5 | 100 | 100 | Y | 8455 | 8505 | 3 | 13 | 48.7 | N | 8265 | 8277 | 4 | 44.5 | 92.6 | X |
| 9100 | 9113 | 4 | 100 | 100 | Y | 8461 | 8498 | 11 | 28.5 | 62.6 | N | 8288 | 8339 | 6 | 39.5 | 1.1 | X |
| 9141 | 9170 | 4 | 76 | 70.6 | Y | 8530 | 8631 | 2 | 4 | 74.1 | N | 8282 | 8348 | 5 | 35.5 | 1.5 | X |
| 9145 | 9165 | 5 | 76 | 70.7 | Y | 8533 | 8628 | 10 | 16.5 | 89.3 | N | 8424 | 8435 | 3 | 14.5 | 17 | X |
| 382 | 537 | 4 | 6 | 40.5 | N | 8546 | 8617 | 4 | 8 | 54.9 | N | 8418 | 8442 | 4 | 28 | 12.7 | X |
| 547 | 565 | 4 | 37.5 | 58.5 | N | 8551 | 8565 | 3 | 6.5 | 60.4 | N | 8469 | 8478 | 2 | 14 | 6 | X |
| 760 | 1010 | 4 | 15 | 75.7 | N | 8578 | 8597 | 3 | 14.5 | 88.7 | N | 8465 | 8482 | 3 | 6.5 | 3 | X |
| 765 | 1005 | 2 | 15 | 75.5 | N | 8840 | 8935 | 2 | 1.5 | 0.2 | N | 8408 | 8493 | 6 | 10 | 5.7 | X |
| 769 | 1001 | 5 | 16 | 75.7 | N | 8845 | 8932 | 4 | 13 | 4.8 | N | 8406 | 8496 | 2 | 8.5 | 5.7 | X |
| 855 | 994 | 2 | 1.5 | 39.1 | N | 8949 | 8961 | 3 | 55 | 95.3 | N | 8404 | 8499 | 2 | 8.5 | 5.7 | X |
| 915 | 970 | 3 | 4 | 20.2 | N | 9011 | 9139 | 5 | 6.5 | 0 | N | 8398 | 8505 | 4 | 9 | 5.7 | X |
| 979 | 991 | 5 | 34.5 | 65.3 | N | 9020 | 9037 | 4 | 6.5 | 0 | N | 8392 | 8511 | 2 | 9 | 5.1 | X |
| 1026 | 1068 | 2 | 42 | 80 | N | 9025 | 9033 | 2 | 6.5 | 0 | N | 8370 | 8534 | 8 | 32 | 11 | X |
| 1200 | 1206 | 2 | 24.5 | 28.5 | N | 187 | 197 | 3 | 43 | 98.8 | X | 8367 | 8537 | 2 | 34.5 | 9.3 | X |
| 1341 | 1795 | 2 | 1.5 | 0.1 | N | 383 | 393 | 2 | 18.5 | 25.9 | X | 8358 | 8546 | 8 | 45 | 10.3 | X |
| 1346 | 1790 | 4 | 1 | 0 | N | 380 | 396 | 2 | 14.5 | 17.6 | X | 8228 | 8551 | 5 | 26 | 1.8 | X |
| 1729 | 1757 | 6 | 15 | 18.3 | N | 533 | 549 | 4 | 31 | 38.2 | X | 8224 | 8555 | 3 | 26 | 1.7 | X |
| 1737 | 1749 | 3 | 26.5 | 21.4 | N | 972 | 979 | 2 | 17.5 | 4.6 | X | 8596 | 8619 | 2 | 22 | 0.4 | X |
| 1796 | 1946 | 7 | 23.5 | 0.8 | N | 968 | 983 | 3 | 17.5 | 4.4 | X | 8591 | 8624 | 2 | 18 | 0.6 | X |

**Figure S1.** *RNAstructure* secondary structure models for a benchmark of six structured RNAs. Cyan lines mark incorrect base pairs; orange lines mark crystallographic base pairs missing in each model. (Figure is in two parts.)

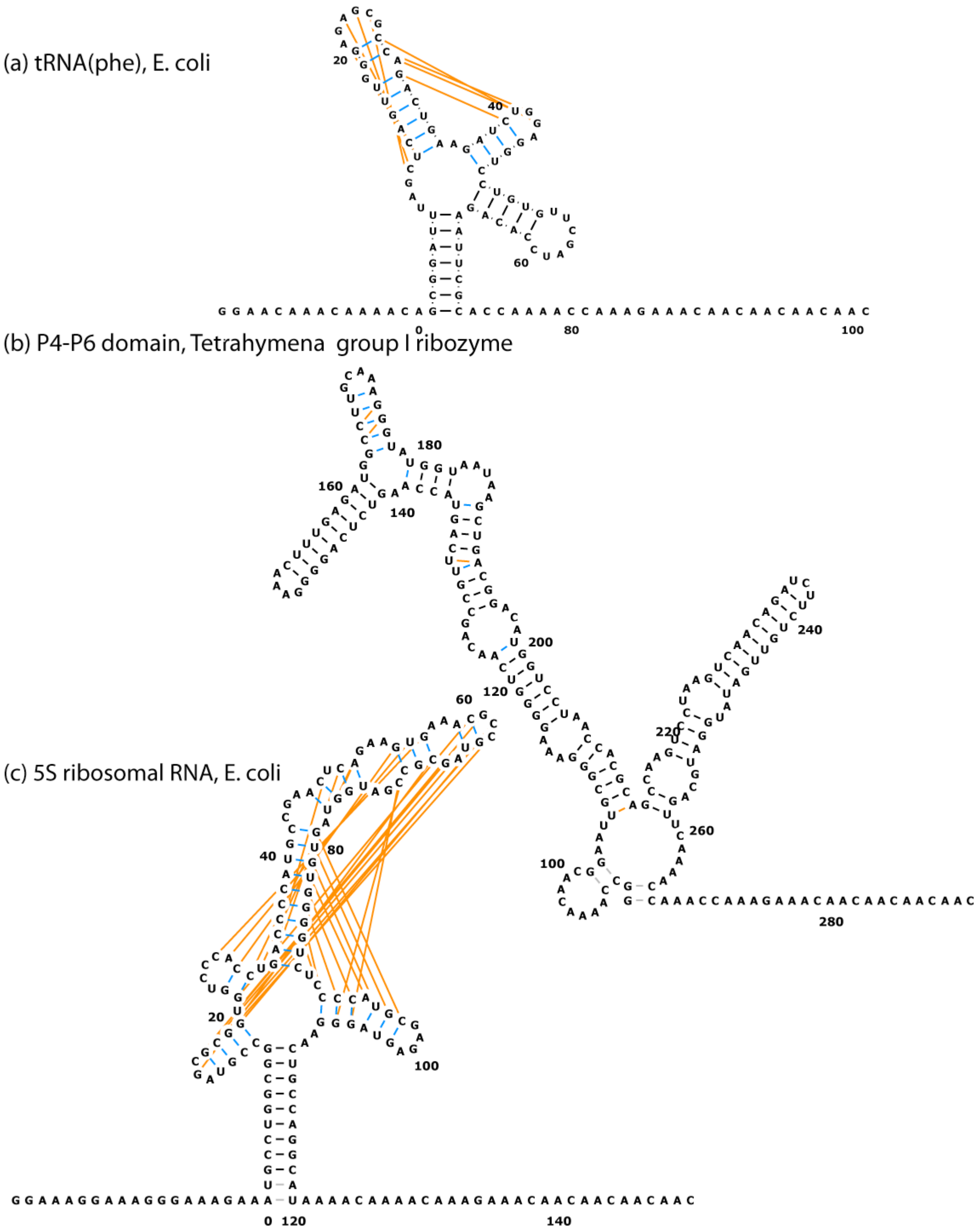

(d) adenosine riboswitch, V. vulnificus

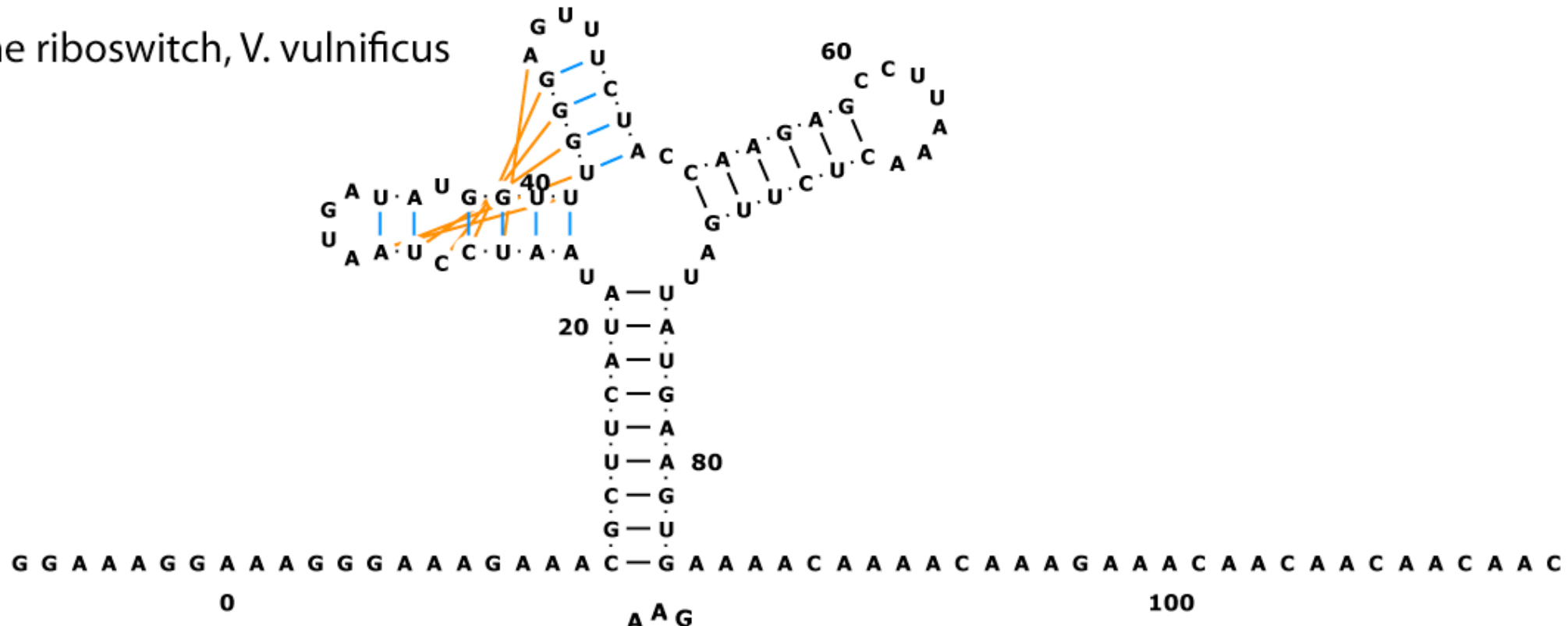

(e) cyclic diGMP riboswitch, V. cholerae

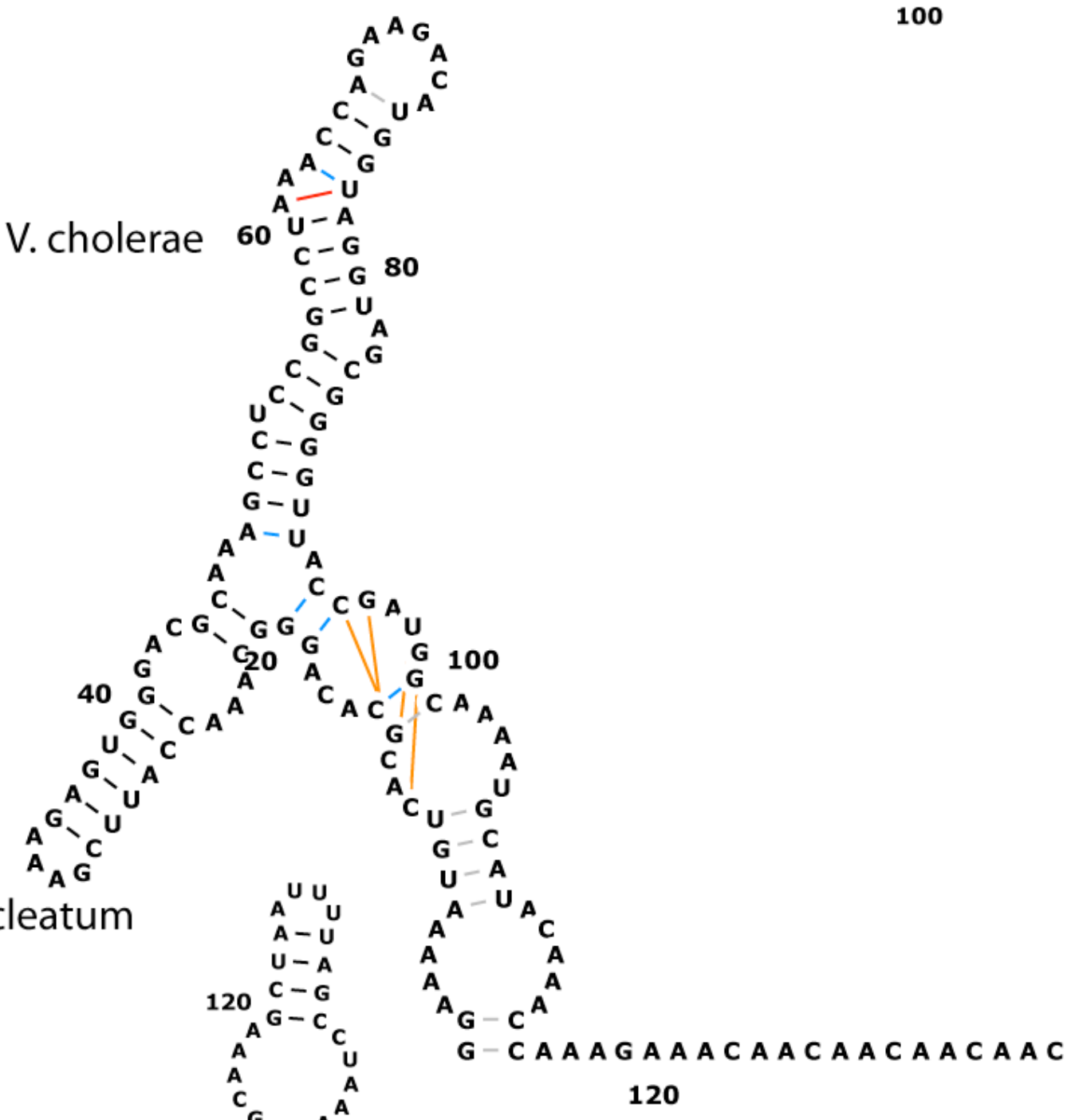

(f) glycine riboswitch, F. nucleatum

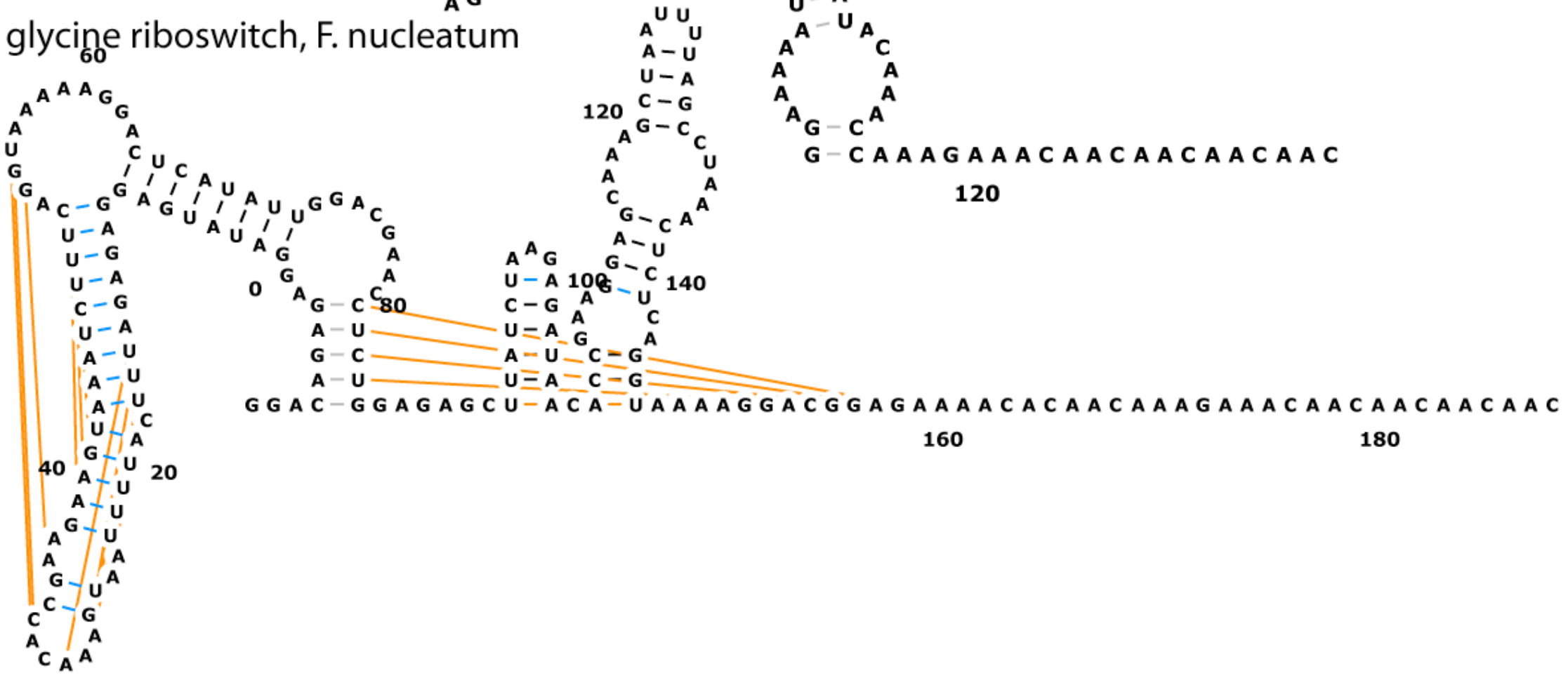

**Figure S2. Recovery of SHAPE-directed model for a previously studied HCV RNA.** As a control, SHAPE data were collected herein for the hepatitis C virus internal ribosomal entry site (HCV IRES) domain II; the resulting SHAPE-directed model agrees with prior work (*1*) and phylogenetic and NMR analyses [see, e.g., (*6*)]. Model is colored by SHAPE reactivity (see color scalebar). Helix confidence estimates from bootstrap analyses (see main text) are given as red percentage values. Flanking sequences (similar to those added to benchmark RNAs) are shown in gray.

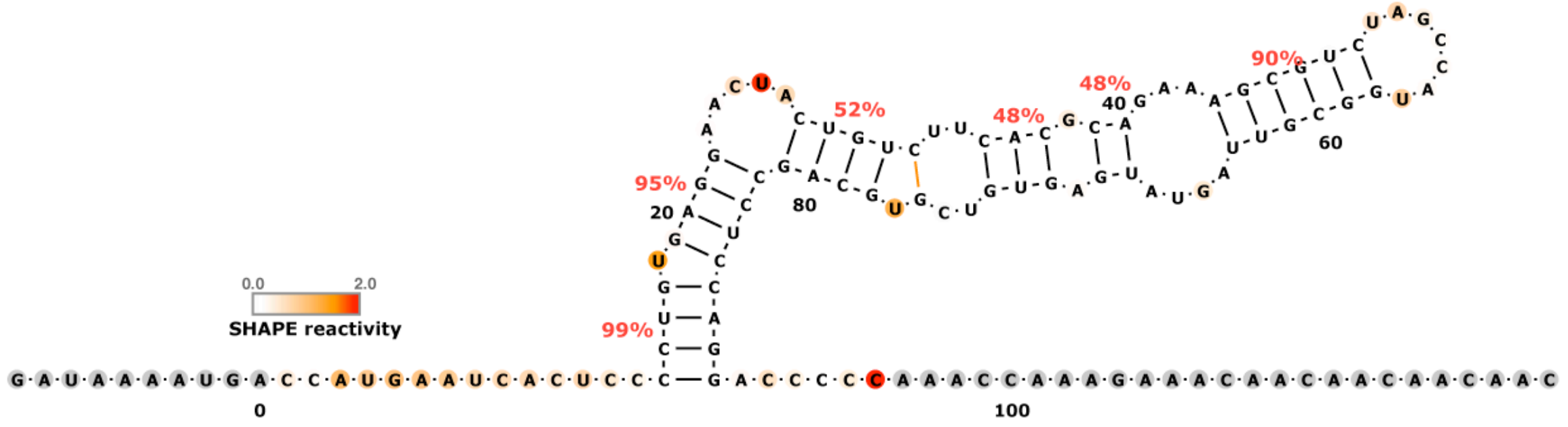

**Figure S3. SHAPE data acquired with different dNTP mix for primer extension, refolding prior to chemical modification, and different DMSO backgrounds.** Colored error bars and lines give background-subtracted data for tRNA$^{phe}$ (*E. coli*) from six experiments: two experiments in which the dATP, dCTP, dITP, and dTTP were used for reverse transcription of modified RNA; and four experiments in which standard dATP, dCTP, dGTP, and dTTP were used. Each experiment involved at least two replicate measurements; error bars represent standard deviations within each experiment. Arrows mark high-variance bands at C nucleotides in dITP experiments (red) due to poor incorporation of dITP, and near G nucleotides in dGTP experiments (blue) due to band compression. 'Refold' experiment 5 (green) involved incubation of RNA at 10 mM $MgCl_2$, 10 mM Na-MES, pH 6.0 at 50 °C for thirty minutes and gave reactivities indistinguishable from conditions without incubation. Low DMSO experiment 6 (dark green) contained 10% DMSO during chemical modification and gave reactivities indistinguishable from conditions used for other experiments (25% DMSO). Black error bars and lines gives the final averaged SHAPE reactivity averaged over all experiments, taking into account the estimated errors (see SI Methods).

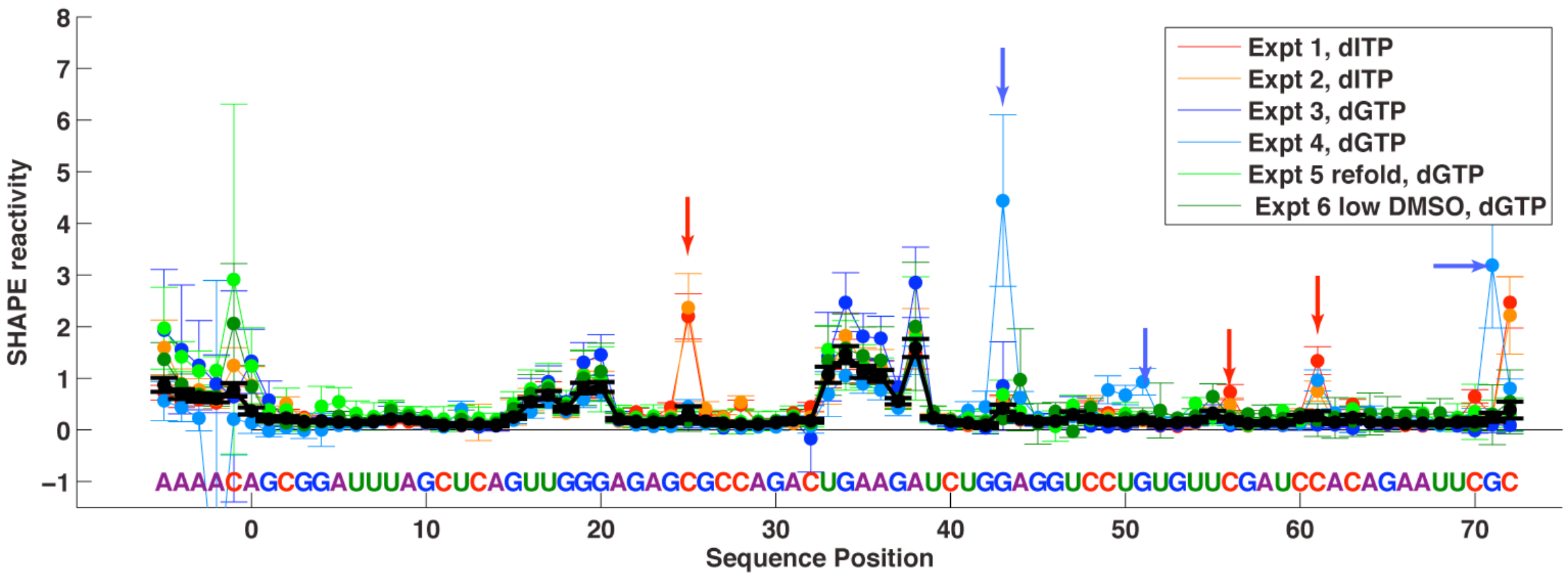

**Figure S4. Demonstration that solution SHAPE data reflect folded or ligand-bound conformations.** Significant differences were observed upon addition of 10 mM $MgCl_2$ with a background of 50 mM Na-HEPES (for the P4-P6 domain & 5S rRNA) and upon addition of ligand with a background of 10 mM $MgCl_2$, 50 mM Na-HEPES, pH 8.0 (for the ligand-binding domains of riboswitches for adenine, c-di-GMP, and glycine). Regions that become protected upon $Mg^{2+}$-induced tertiary folding or ligand binding are annotated on the data, and compare well to expectations from previous biophysical and crystallographic studies (*7-13*).

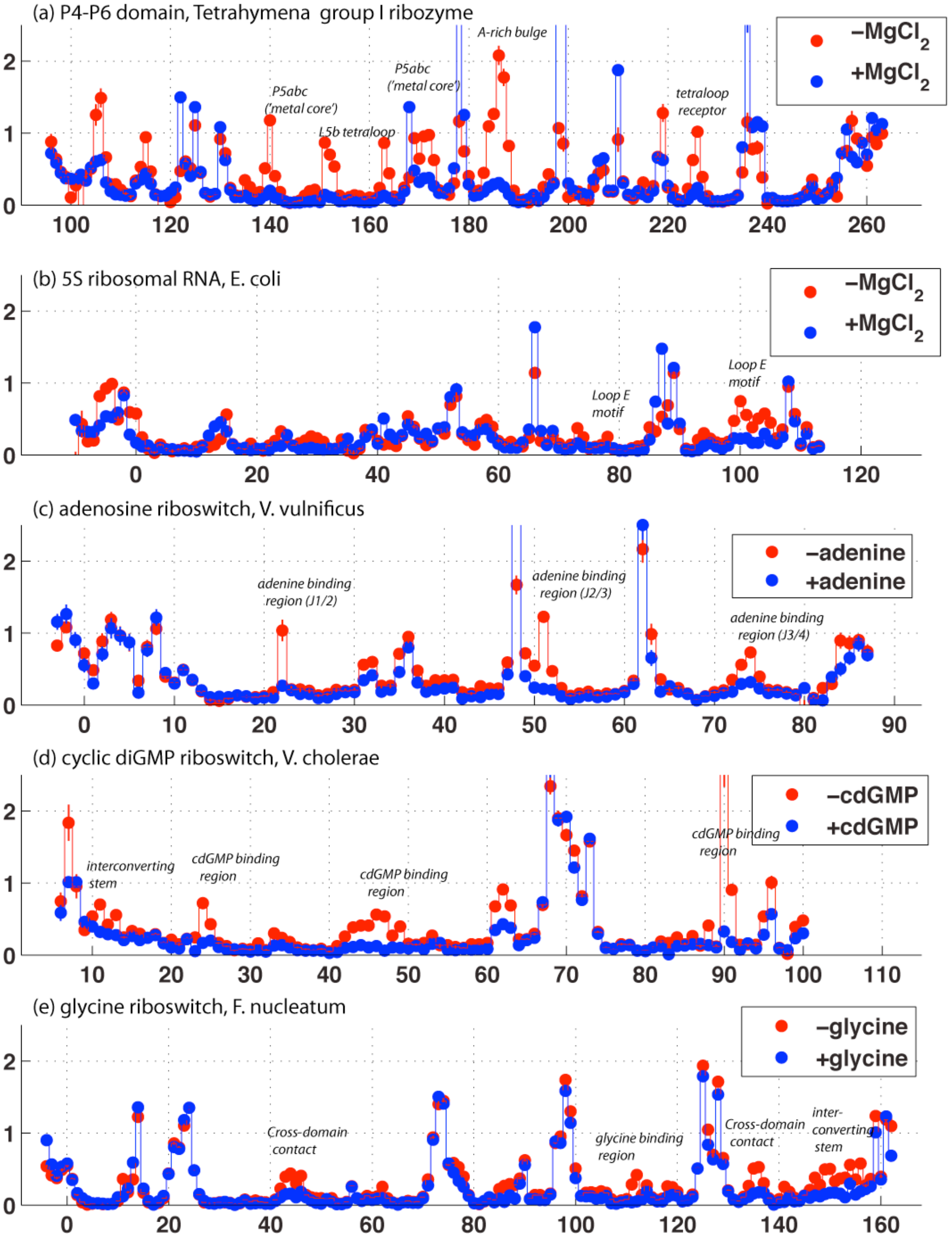

**Figure S5. Partition-function and bootstrap analysis of SHAPE-directed secondary structure models**. A confidence estimate for each helix in each of the six benchmark SHAPE-directed models was determined by (1) partition-function-based Boltzmann probabilities and (2) a nonparametric bootstrap analysis (repeating the modeling with 'replicate' data sets generated by randomly resampling the data with replacement). The confidence estimates for the two analyses correlate approximately, but partition function probabilities are skewed to higher values than bootstrap probabilities. Helices that agree (blue) or disagree (red) with crystallographic secondary structures are plotted separately. Dashed lines mark 80% and 55% separatrix values, above which two incorrect helices are observed, and 29 and 31 correct helices are observed for Boltzmann probability and bootstrap analyses, respectively.

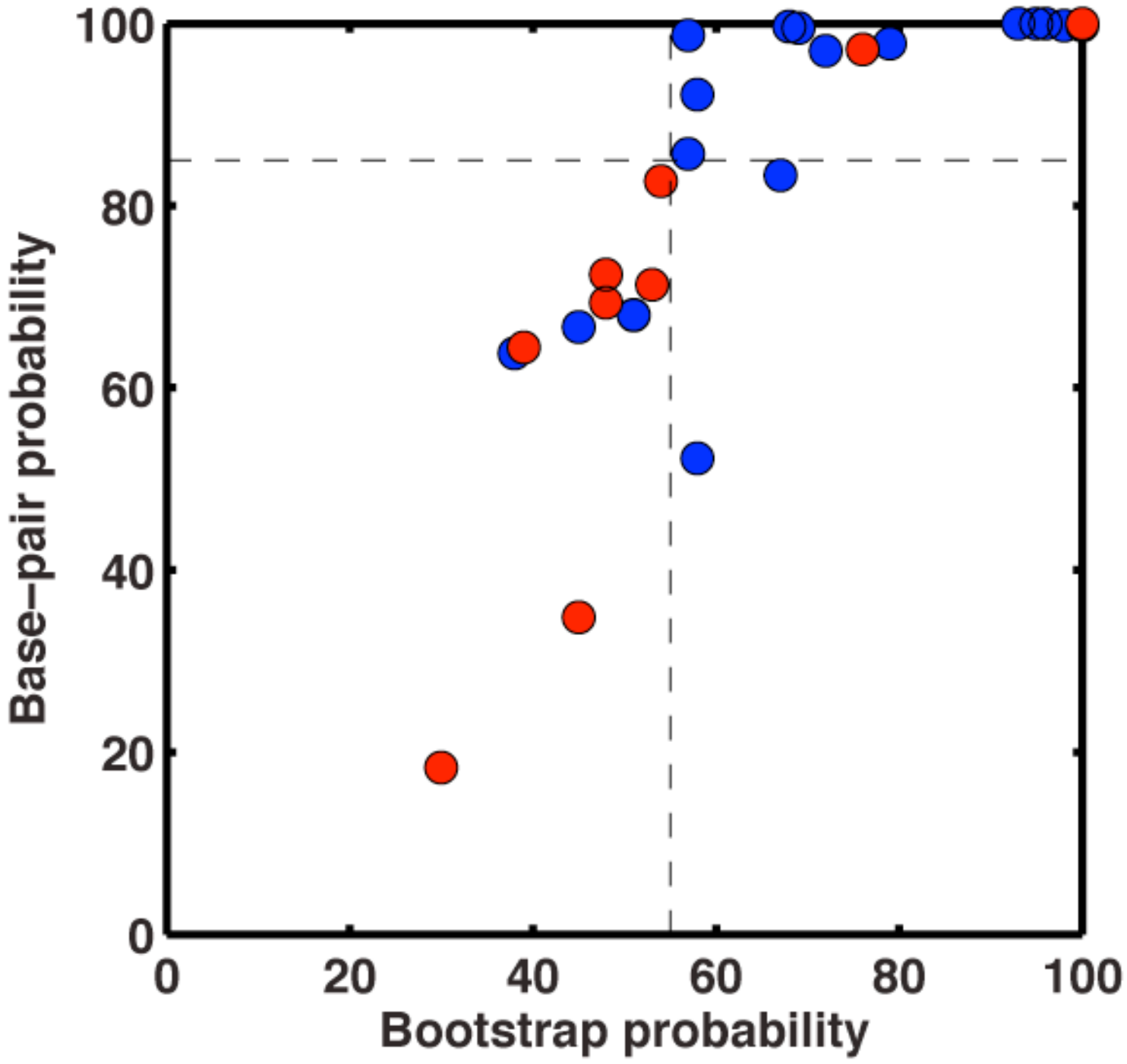

**Figure S6. Sensitivity of minimum-energy model and robustness of bootstrap values to small changes in tRNA SHAPE data.** (a) Comparison of SHAPE data sets obtained by averaging over all collected data (14 replicates; blue) and by averaging over just those data collected with primer extension with standard deoxynucleotide triphosphates (no dITP; i.e., dATP, dCTP, dGTP, dTTP) (6 replicates; red). (b,c) Minimum-energy SHAPE-directed secondary structures are different for the two data sets in the pairings of the third helix; bootstrap values given as red percentage values. (d,e) Helix probabilities from bootstrap analysis shown in grayscale, with 0 to 100% shown as white to black. Bootstrap values at alternate locations of third helix are shown as red percentage values; they are similar for the two data sets. Red squares mark crystallographic base pairs.

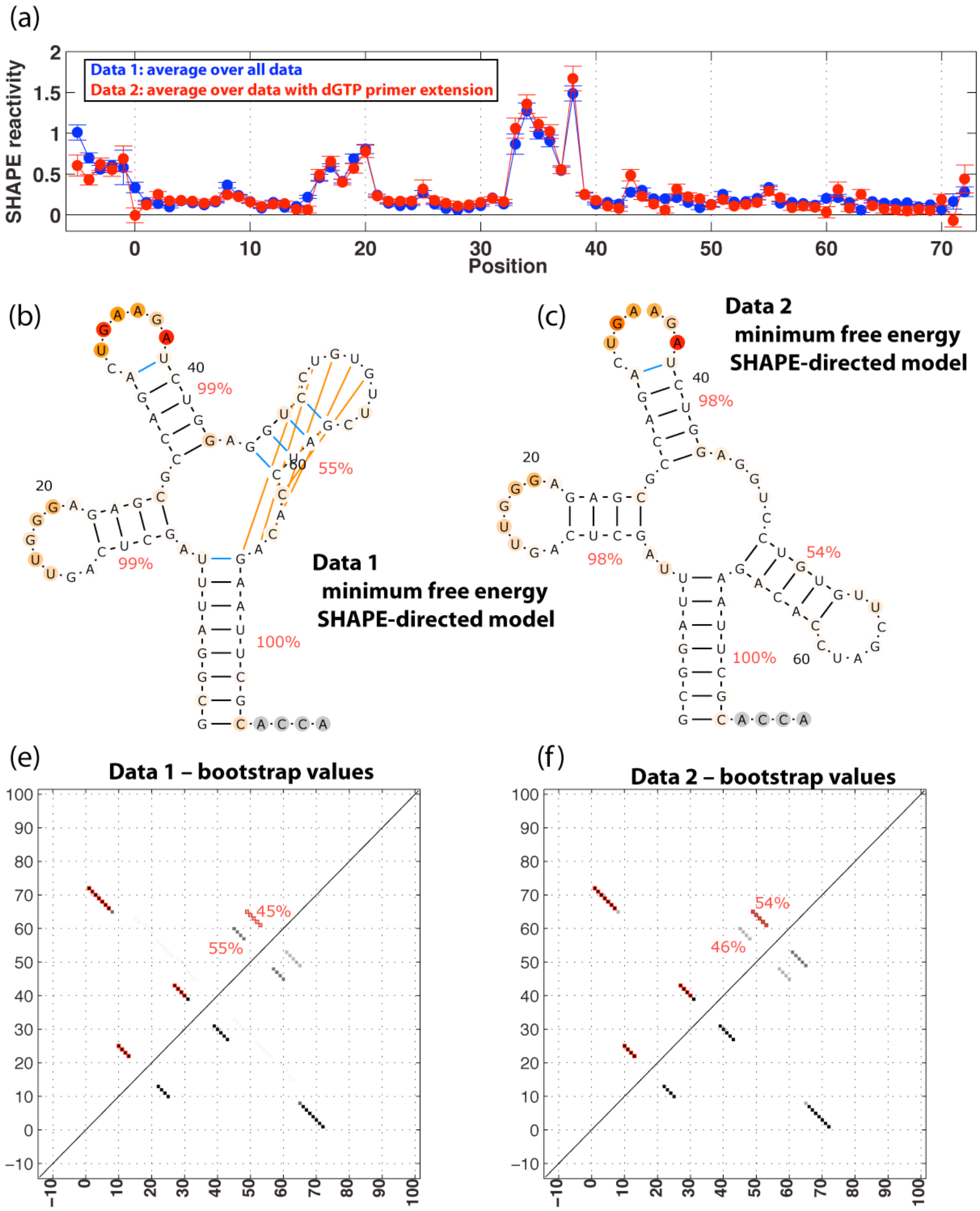

**Figure S7. HIV-1 secondary structure helix confidence values compared to SHAPE reactivities.** For each helix in the HIV-1 secondary structure model (*5*), the median SHAPE reactivity for nucleotides in the helix was computed, and plotted against bootstrap values. Blue line marks median reactivity over all nucleotides. High-bootstrap-value helices (e.g., four helices in TAR, three helices in gag-pol, and the gp120 signal-peptide stem; shown as red x's) typically have low median SHAPE reactivities. However, the converse is not true. Low-reactivity helices frequently have poor bootstrap values, indicating the existence of multiple secondary structures consistent with the data while still protecting the associated regions.

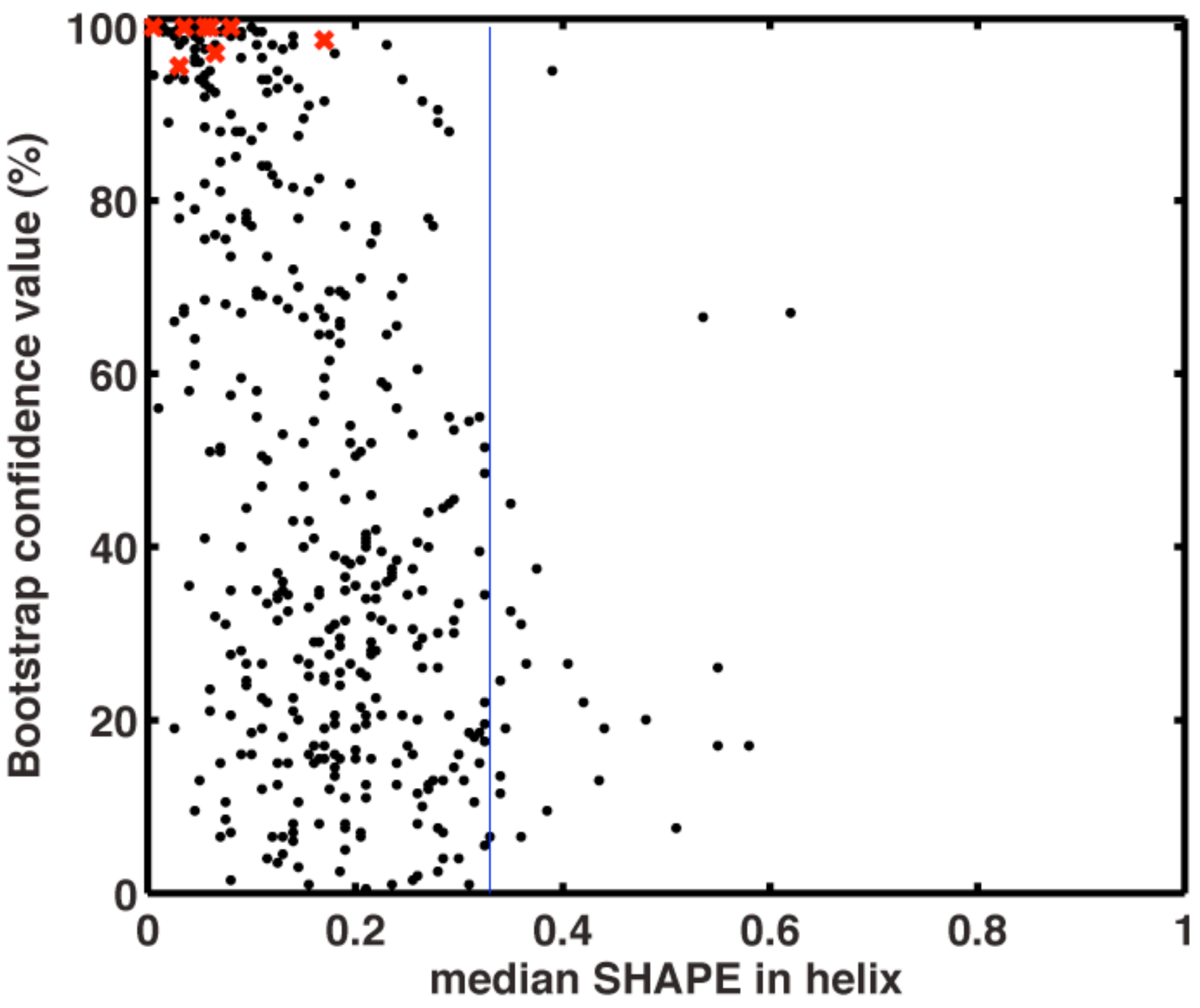

**Figure S8. Histogram and fit of SHAPE reactivities**. SHAPE reactivities of all residues for the six test RNAs (black; see SI Table S1), compared to least-squares fit (red) to a simple probability distribution P($x$). The distribution was assumed to take the form P( $x$ ) = $N$ exp( $F_+|x-x_0|$ ) for $x > x_0$; and P($x$) = $N$ exp( $F_-|x-x_0|$ ) for $x \leq x_0$. The presented fit is for $x_0 = 0.06$; $F_+ = 5.0$; $F_- = 25.0$.

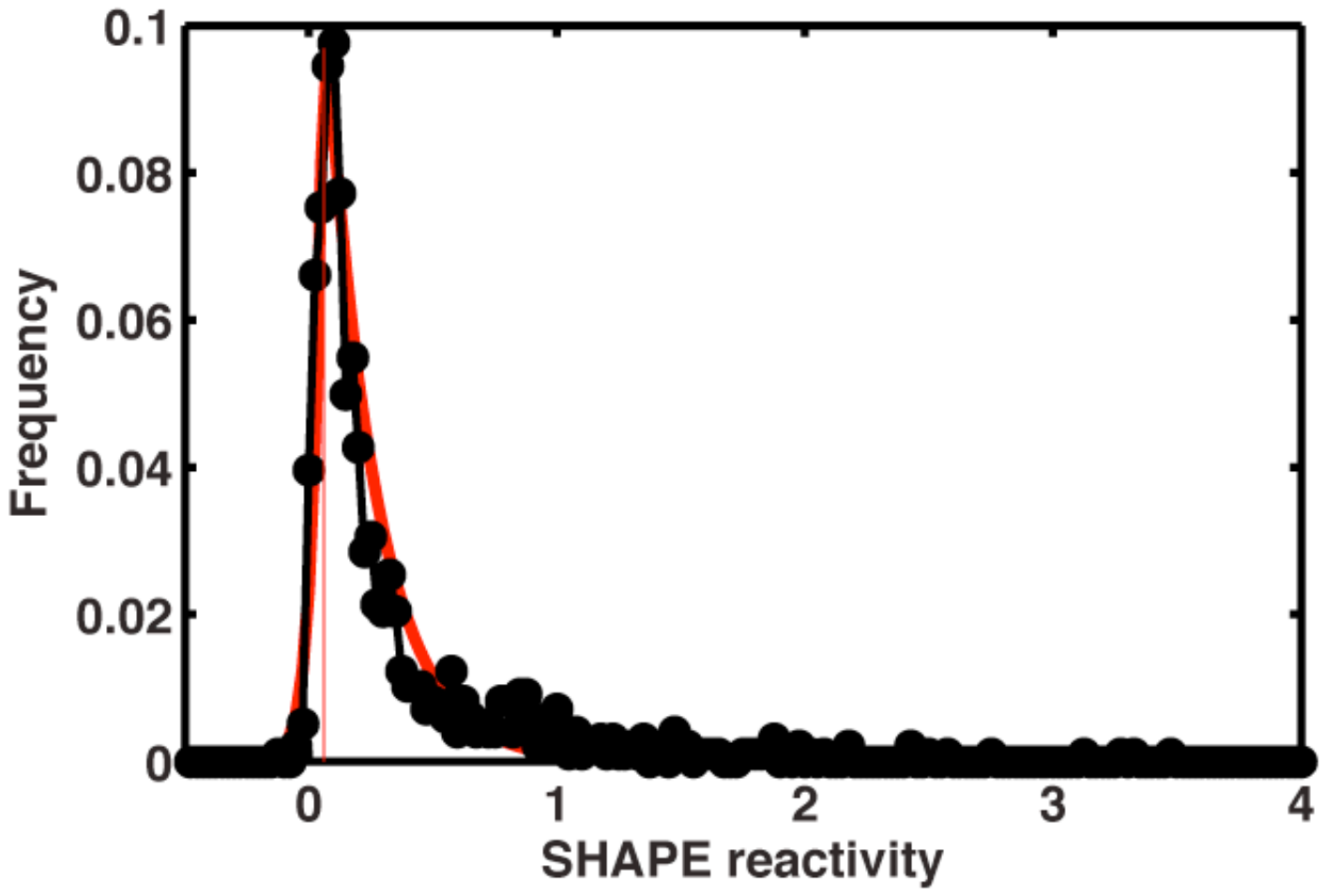

**References for Supporting Information**